\documentclass[useAMS,usenatbib]{mnras}
\usepackage[utf8]{inputenc}
\usepackage{amsmath}
\usepackage{amsfonts}
\usepackage{amssymb}
\usepackage{graphicx}
\usepackage{epstopdf}
\usepackage{listings}
\usepackage{multirow}

\usepackage{tikz}
\usetikzlibrary{shapes.geometric, arrows}
\tikzstyle{input} = [trapezium, trapezium left angle=70, trapezium right angle=110, minimum width=1cm, minimum height=1cm, text centered, text width=0.5cm, draw=black, fill=white!30]
\tikzstyle{process} = [rectangle, minimum width=1cm, minimum height=1cm, text centered, text width=2.5cm, draw=black, fill=white!30]
\tikzstyle{decision} = [diamond, minimum width=1cm, minimum height=1cm, text centered, text width=2.5cm, aspect=2.5, draw=black, fill=white!30]
\tikzstyle{end} = [rectangle,rounded corners,minimum width=2.5cm, minimum height=1cm,text centered, draw=black, fill=white!30]
\tikzstyle{arrow} = [thick,->,>=stealth]

\usepackage{color}
\usepackage{xcolor}

\title{Predicting interstellar radiation fields from chemical evolution models}

\author[M.~Romero et al.]
{M.~Romero$^{1}$, P. Corcho-Caballero$^{1,2}$, I. Millán-Irigoyen$^{3}$,  M.~Moll\'{a}$^{3}$, and 
\newauthor
Y.~Ascasibar$^{1}$ \\
$^{1}$ Departamento de F\'{i}sica Te\'{o}rica, Universidad Aut\'{o}noma de Madrid, E-28049 Madrid, Spain\\
$^{2}$ Australian Astronomical Optics, Macquarie University, 105 Delhi Rd, North Ryde, NSW 2113, Australia\\
$^{3}$ Departamento de Investigaci\'{o}n B\'{a}sica, CIEMAT, Avda. Complutense 40. E-28040 Madrid, Spain\\
}

\pagerange{\pageref{firstpage}--\pageref{lastpage}}
\pubyear{2022}

\newcommand{\dd}{\mathrm{d}}

\newcommand{\cloudyCommand}[1]{\texttt{#1}}

\newcommand{\skirtCommand}[1]{\texttt{#1}}

\begin{document}

\maketitle
\label{firstpage}

\begin{abstract}
We present a self-consistent prediction of the interstellar radiation field (ISRF), from the extreme ultraviolet (EUV) to sub-mm range, based on two chemical evolution models of a Milky Way-like galaxy (MWG). To this end, we develop a new tool called {\sc Mixclask} to include gas emission, absorption and scattering from the photoionization code {\sc Cloudy} into the Monte Carlo radiative transfer code {\sc Skirt}.
Both algorithms are invoked iteratively, until the physical properties of the ISM converge.
We have designed a first test, reminiscent of a H{\sc ii} region, and we find that the results of {\sc Mixclask} are in good agreement with a spherically symmetric {\sc Cloudy} simulation.
Both MWG models based on chemical evolution codes give results broadly consistent with previous empirical models reported in the literature for the ISRF of our Galaxy, albeit they systematically underestimate the mid-infrared emission.
We also find significant differences between our two models in the whole ultraviolet range, not fully explored in previous ISRF models.
These results show the feasibility of our method of combining radiative transfer with chemical evolution models: there is increased predictive power and the interstellar radiation field obtained provides further constraints on the model parameters.
Python source code to implement our method is publicly available at \url{https://github.com/MarioRomeroC/Mixclask}.
\end{abstract}


\section{Introduction}
\label{sec_Introduction}

Galaxies are complex systems, whose evolution is governed by multiple interactions between their main constituents.
One means to investigate the problem is by chemical evolution models, hereinafter CEM, where the interstellar medium (ISM) is considered to be composed of gas in different phases (e.g. ionized, atomic and molecular) and/or chemical abundances, as well as solid dust particles.
A CEM follows how gas is accreted into the galaxy from the intergalactic medium and cools down, going through the different phases, until it eventually forms stars.
When these die, they return a fraction of their mass to the ISM in the form of both gas (enriched with newly synthesised chemical elements) and dust.
They also inject a significant amount of kinetic and thermal energy, collectively known as feedback, that plays an important role in regulating the physical properties of the ISM and the conversion rate of gas and dust into new generations of stars. 
The problem is solved semi-analytically (numerically integrating a system of analytic differential equations), which is extremely efficient from a computational point of view, and thus CEM provide a useful tool to test new hypotheses about the different ingredients that play a role in galaxy formation and evolution~\citep[see e.g.][]{Ascasibar+15, Molla+2015, Molla+17, Millan-Irigoyen+20}.

Here we present an iterative scheme to self-consistently predict the interstellar radiation field (ISRF) of a galaxy based on the results of a chemical evolution model.
The absorption and emission of photons is one of the most efficient means to transfer energy within the galaxy and radiating it away from the system towards the intergalactic medium. 
It is well known since a long time ago \citep[e.g.][]{Habing68, Mathis+83, Black87} that modelling the ISRF is critical in order to understand the intricate network of processes that set the physical state of the ISM. 
In the absence of an electromagnetic radiation field, gas cooling may be accounted for under the simplifying approximation of collisional ionization equilibrium, which yields a cooling function that only depends on the gas temperature, density and chemical composition~\citep[see e.g.][]{SutherlandDopita93}.
However, the ISRF itself may have a significant effect on the radiative cooling and heating curves~\citep[e.g.][]{Wiersma+09, GnedinHollon12, Sarkar+21b, Robinson+21}, and therefore on its thermal energy balance as well as on the the equilibrium between the ionized, atomic, and molecular phases~\citep[e.g.][]{Obreja+19}.
This, in turn, has profound implications on all the processes that take place within the galaxy, including star formation and feedback, or the growth and destruction of dust grains.
It has been shown, for example, that the shielding of the surrounding radiation field affects the outer shell structure of supernova remnants~\citep{Romero+21}, as well as the propagation and energy losses of cosmic rays~\citep[e.g.][]{DRAGON2,Porter+17}.

Thus, modelling the ISRF is extremely relevant from the point of view of galaxy formation and evolution.
Unfortunately, solving the radiative transfer equation to compute the intensity of the radiation field as a function of wavelength, spatial location within the galaxy, and direction in the sky is a formidable and computationally very expensive challenge.
Multiple methods have been developed in the literature, every one with its own strengths and weaknesses, to tackle different aspects of the problem.
Some of them, like {\sc Cloudy}~\citep{Cloudy17} or {\sc mappings}~\citep{MappingsIII, MappingsV, Jin+22}, focus on the ISM chemistry and thermodynamics, thoroughly accounting for its physical properties, but they are limited to radiative transfer in one dimension (cylindrical or plane-parallel symmetry).
Others, such as {\sc mocassin}~\citep{Mocassin, Mocassin2}, {\sc sunrise}~\citep{Sunrise}, {\sc Skirt}~\citep{SkirtOrigins, Skirt9}, {\sc hyperion}~\citep{Hyperion}, {\sc radmc3d}~\citep{Radmc3d} or {\sc torus}~\citep{Harries+19}, specialise in solving the radiative transfer equation in three spatial dimensions by means of Monte Carlo techniques, using different approximations to model the physical properties of both the stellar population and the interstellar medium.
Finally, there are also analytical works that compute the ISRF as a sum of blackbodies with different properties~\citep[e.g.:][]{Mathis+83,Delahaye+10,WechakamaAscasibar14} or by adding radiation sources, using a probability distribution for each luminosity, extinguished by dust~\citep[e.g.][]{Bialy20}.

Previous attempts to account for the effects of the ISRF in the context of CEM~\citep[e.g.][]{Millan-Irigoyen+20} and numerical hydrodynamical simulations~\citep[e.g.][]{Gnedin&Abel01, Petkova&Springel09, Wise&Abel11, Rosdahl+13, Kannan+19, trevr, treeray,Petkova+21} focus on the number of photons that can ionize (or dissociate) the atomic (molecular) gas, but solving the full radiative transfer problem, coupled with chemical evolution and hydrodynamics in a cosmological context, is still beyond current technical capabilities. 
This way, synthetic observations in different photometric bands and/or specific atomic and molecular lines have also been produced by post-processing the results of hydrodynamical simulations with
{\sc mocassin} \citep[e.g.][]{Karczewski+13},
{\sc sunrise} \citep[e.g.][]{Torrey+15,SmithHayward18,Guidi+18},
{\sc hyperion} \citep[e.g.][]{Robitaille+12},
{\sc radmc3d} \citep[e.g.][]{Seifried+20},
{\sc torus} \citep[see][and references therein]{Harries+19},
and
{\sc Skirt} \citep[e.g.][]{Camps+18,Camps+22,Kapoor+21}, among others.


Nevertheless, a realistic prediction of the spectral energy distribution of the ISRF as a function of spatial location, taking into account radiative transfer within a self-consistent model galaxy, has never been attempted.
Here, we propose a new scheme that takes advantage of the photoionisation code {\sc Cloudy} and the Monte Carlo radiative transfer implemented in {\sc Skirt} to compute the ISRF from the output of a chemical evolution model.
The spectrum emitted by the stellar population is obtained from the {\sc PopStar} evolutionary synthesis models~\citep{PopStar2009}, while the emissivities and opacities of gas and dust in the ISM are based on the {\sc Cloudy} results.
The latter, however, depend on the input ISRF, which is iteratively computed by {\sc Skirt} until convergence is reached.
This procedure, akin to the algorithm implemented in other codes that take into account the gas phase of the ISM \citep[such as e.g. {\sc torus}, described in][]{Harries+19}, is essential in order to correctly model thermal and ionisation equilibrium, the absorption and emission coefficients of the ISM, and the ISRF beyond the Lyman limit, all of which are of the utmost importance in order to understand potential feedback effects on the star formation cycle. 

Our method is fully described in Section~\ref{sec_mixcalsk}, where we test it against a simple one-dimensional case with spherical symmetry and compare the results with those obtained by the {\sc Cloudy} code.
Then, the method is applied to the chemical evolution models \citet{Millan-Irigoyen+20} and \citet{Molla+2023}, meant to represent spiral galaxies similar to the Milky Way {in order to illustrate its ability to make a realistic prediction of the ISRF of a model galaxy}.
The results are presented in Section~\ref{sec_science} and compared with previous efforts in the literature, which characterise the Galactic ISRF by fitting the intensity observed from the Earth at different wavelengths to a phenomenological description of the spatial distribution of radiation sources and sinks \citep[e.g.][]{Mathis+83, Black87, PorterStrong06, Delahaye+10, Popescu+17}.
This is, to our knowledge, the first time that a self-consistent CEM is used to predict the ISRF of our Galaxy, from the extreme ultraviolet (EUV, {above $13,6$~eV}, barely discussed in the literature)
to the sub-mm range, as a function of spatial location.
Our main conclusions and future prospects are briefly summarised in Section~\ref{sec_Conclusions}.

\section{Radiative transfer}
\label{sec_mixcalsk}

\subsection{Mixing Cloudy and Skirt}
\label{sec_flowchart}

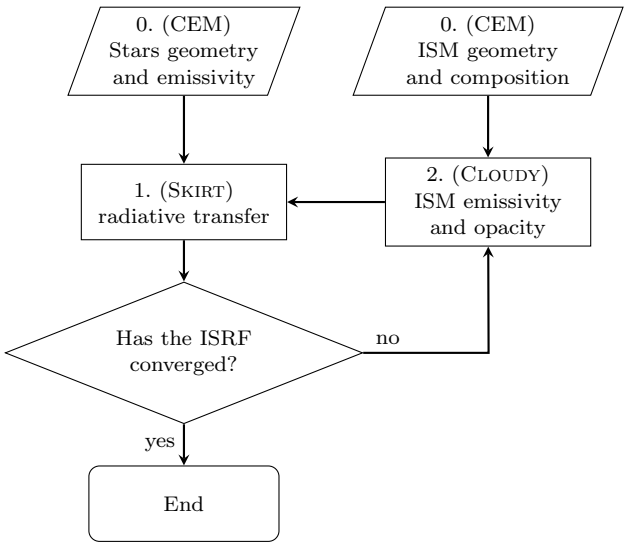
\begin{figure}
\begin{tikzpicture}[node distance=2cm]
    \node (stars) [input, text width=2cm] {0. (CEM) \\ Stars  geometry and emissivity};
    \node (gas) [input, right of=stars, xshift=2cm, text width=2.5cm] {0. (CEM) \\ ISM geometry and composition};
    \node (skirt) [process, below of=stars] {1. ({\sc Skirt}) \\ radiative transfer};
    \node (cloudy) [process, right of=skirt, xshift=2cm] {2. ({\sc Cloudy}) \\ ISM emissivity and opacity};
    \node (conv) [decision, below of=skirt] {Has the ISRF converged?};
    \node (finish) [end, below of=conv] {End};
    \node (control) [right of=conv, xshift=1.9cm]{};
    \draw [arrow] (stars) -- (skirt);
    \draw [arrow] (gas) -- (cloudy);
    \draw [arrow] (skirt) -- (conv);
    \draw [arrow] (cloudy) -- (skirt);
    \draw [arrow] (conv) -- node[anchor=east]{yes} (finish);
    \draw [arrow] (conv) -- node[anchor=south,xshift=-0.5cm]{no} (control.east) -- (cloudy);
\end{tikzpicture}
\caption{{\sc Mixclask} flowchart. The stellar spectra and ISM (i.e. gas and dust) composition are used as input, and {\sc Mixclask} iteratively predicts the ISRF, updating the emissivity and opacity of the ISM until convergence is achieved (see text).}
\label{Fig_flowchart}
\end{figure}

\begin{table*}
\renewcommand{\arraystretch}{1.5}
\begin{tabular}{c | c | l}
    Input & Options/parameters & Description \\
    \hline\hline
    Stellar geometry
     & {point} & Point source at the origin. \\
     & {shell}: $\{R_{\rm in},R_{\rm out} \}$ & Inner and outer radius of a spherical shell. \\
     & {ring}: $\{R, w, h \}$ & Middle radius, half-width, and height of a cylindrical ring. \\
    Stellar luminosity & $\lambda L_{\lambda}$ & Luminosity emitted by the stars in the region. \\
    \hline
    ISM geometry
     & {shell}: $\{R_{\rm in},R_{\rm out} \}$ & Inner and outer radius of a spherical shell. \\
     & {ring}: $\{R, w, h \}$ & Middle radius, half-width, and height of a cylindrical ring. \\
    ISM mass & $M$ & Total mass of gas and dust in the region. \\
    Number density & $n_H$ & Hydrogen particles (ions, atoms, and molecules) per unit volume. \\
    Chemical composition
     & -- & Solar abundances \citep{GASS}. \\
     & $X_i = {M_i}/{M_{\rm gas}}$ & (optional) Mass fractions of elements $i$ being from He to Zn. \\
     & $Y=X_{\rm He}$, $Z = \sum_{i={\rm Li}}^{\rm Zn} X_i $ & (optional) Helium mass fraction and metallicity. \\
    Dust content
     & -- & Default ISM and PAH composition (see Cloudy documentation) \\
     & $DTG = M_{\rm dust}/M_{\rm gas}$ & (optional) Dust (grains+PAH) to gas ratio. \\
     & $q_{\rm PAH} = M_{\rm PAH}/M_{\rm dust}$ & (optional) PAH to dust ratio. \\
    \hline
\end{tabular}
\caption{Input parameters accepted by {\sc Mixclask} for each modelled region in the simulated domain.
}
\label{Table_inputs}
\renewcommand{\arraystretch}{1.0}
\end{table*}

\begin{table*}
\renewcommand{\arraystretch}{1.5}
\begin{tabular}{c | c | l}
    Output & Notation & Description \\
    \hline\hline
    ISRF
     & $4\pi \lambda J_{\lambda}$ & Mean intensity at the centre of each ISM region. \\
     &  & (optional) Mean intensity at other, user-defined locations. \\
    ISM optical properties
     & $\lambda L_{\lambda}$ & Luminosity emitted by the gas and dust in each region. \\
     & $\kappa$, $\varpi$ & Mass extinction coefficient and albedo. \\
    \hline
\end{tabular}
\caption{Relevant output generated at every {\sc Mixclask} iteration.
}
\label{Table_output}
\renewcommand{\arraystretch}{1.0}
\end{table*}

We have developed a Python code named {\sc Mixclask}\footnote{\url{https://github.com/MarioRomeroC/Mixclask}} to calculate the radiation field over a certain domain filled with stars and ISM (gas and dust) material.
This is achieved by combining {\sc Skirt} \citep{Skirt9} to perform radiative transfer and {\sc Cloudy} \citep{Cloudy17} to compute the emissivity and opacity of the ISM.
Further technical details about the configuration of both codes are given in Appendix~\ref{app_Technical}.

The flowchart of our method is summarised in Figure~\ref{Fig_flowchart}.
All {\sc Mixclask} simulations are performed in two dimensions with axial symmetry, and the cylindrical computational domain is divided into several separate regions containing stars and ISM material, that may in turn adopt different predefined geometries.
The total physical extent of the simulated domain is delimited by the farthest regions from the coordinate origin.
When the model regions are defined as spherical shells, the largest $R_{out}$ sets the radial and vertical extent of the cylindrical simulation domain.
When ring regions are specified, the radial extent includes the radius and width $R+w$ of the outermost ring.
For the vertical direction, on the other hand, the simulated domain is by default ten times thicker than the maximum characteristic scale height $h$ of the disk (the exact factor may be tuned by the user) in order to allow for the computation of the ISRF far from the mid-plane.
{\sc Mixclask} allows to mix both geometries in a single simulation.
In that case, the simulation boundaries are selected with the farthest distances.

The input physical properties of stellar and ISM regions are quoted in Table~\ref{Table_inputs}.
For the stellar component, it is necessary to define their geometry and spectra, given in terms of the total luminosity
\begin{equation}
    \lambda L_{\lambda} = \lambda \int{j_{\lambda}\ \dd V \dd\Omega},
    \label{eq_luminosity}
\end{equation}
where $j_{\lambda}$ denotes the specific emission coefficient (energy emitted per unit time, volume, solid angle, and wavelength) of each region.
This information is directly imported by \textsc{Skirt}, and it suffices to fully describe the stellar population as a source term in the radiative transport equation.

In contrast, the matter in the interstellar medium can both emit and absorb photons.
The ISM in {\sc Mixclask} is considered to be a mixture of gas and dust, where the latter is further divided into grains and polycyclic aromatic hydrocarbons (PAH).
The user must define its geometry as a sum of shells and/or rings (that may differ from those describing the stellar component) and specify their chemical compositions.
The minimum set of parameters consists of the ISM mass and Hydrogen number density within each region.
The code will assume the default options in {\sc Cloudy} (see Appendix~\ref{ssec_cloudy}) unless the Helium mass fraction $Y$ and overall metallicity $Z$ (or individual element abundances $X_i$) are specified.
Dust properties are defined with the dust-to-gas (DTG) and PAH-to-dust ($q_{\rm PAH}$) ratios.

The main output of {\sc Mixclask} is summarised in Table~\ref{Table_output}.
The most important quantity is, of course, the mean intensity of the ISRF:
\begin{equation}
    4\pi\lambda J_{\lambda} = \lambda \int{ I_{\lambda}\ \dd\Omega},
    \label{eq_meanIntensity}
\end{equation}
where $I_{\lambda}$ is the radiative energy per unit time, area, solid angle, and wavelength.
The mean intensity is evaluated at the centre of every ISM region, in order to determine their individual emissivities and opacities, as well as at any other location inside the simulation domain specified by the user.

During the first iteration, {\sc Mixclask} computes $4\pi\lambda J_{\lambda}$ by running {\sc Skirt} with stars only, neglecting any emission or absorption by the ISM.
Then, individual {\sc Cloudy} simulations are run for each ISM region using the radiation field obtained in the last step.
Since {\sc Cloudy} distinguishes between the illuminated and shielded faces, whereas the ISRF in our problem is assumed to be homogeneously distributed over each region, we approximate them as a {\sc Cloudy} slabs whose thickness $s$ is equal to their half-widths ($s=w$ for rings and $s=\frac{R_{\rm out}-R_{\rm in}}{2}$ for spherical shells) and simulate the propagation of the average ISRF from the centre of the region ({\sc Cloudy}'s illuminated face) outwards.

{\sc Cloudy} computes the contribution of the ISM to the mean intensity $\tilde J_\lambda$ emerging through the shielded face of each slab, as well as its total optical depth $\tau$ and the scattering contribution $\tau_{\rm sca}$ as a function of wavelength.
From these outputs, which we downsample to the desired set of wavelengths by energy-conserving interpolation, we must obtain the total luminosity emitted by each {\sc Mixclask} ISM region as well as their mean opacity and albedo, which are the quantities required by {\sc Skirt}.

Expressing the average extinction coefficient of the slab as $\langle \alpha \rangle \equiv \tau /s$ in terms of its optical depth $\tau$ and thickness $s$, and making the approximation that
$ \tilde J_\lambda \approx
 \frac{\langle j_\lambda \rangle}{\langle \alpha \rangle}
 \left(1-e^{-\tau }\right)$
in order to account for self-absorption, the total luminosity emitted by the region will be given by:
\begin{equation}
    \lambda L_{\lambda} \equiv 4\pi \lambda \langle j_\lambda \rangle V
    \approx \frac{4\pi \lambda \tilde J_{\lambda}}{s} \frac{\tau}{1-e^{-\tau}} V,
    \label{eq_Lcloudy}
\end{equation}
in terms of the average emission coefficient $\langle j_\lambda \rangle$ based on the {\sc Cloudy} results and the volume $V$ of the region obtained from the ISM mass and density $\rho$ as $V = M/\rho$.
This density is computed from dust-to-gas ratio and the chemical gas abundances provided by {\sc Cloudy} for each region:
\begin{equation}
    \rho = (1+DTG) \sum_{i=\rm H}^{\rm Zn} A_i m_H n_i
\end{equation}
where $A_i$ and $m_H$ are the mass number of the element $i$ and the Hydrogen mass, respectively.

The mass extinction coefficient can be trivially obtained as
\begin{equation}
    \kappa = \frac{\tau}{\rho\,s},
    \label{eq_extinctionCoef}
\end{equation}
and the albedo is calculated as the ratio
\begin{equation}
    \varpi = \frac{\tau_{\rm sca}}{\tau}
    \label{eq_albedo}
\end{equation}
between the scattering and optical depth as a function of wavelength.

\subsection{Convergence and statistical uncertainties}
\label{ssec_ConvergenceAndErrors}

Once the optical properties of the ISM are updated according to expressions~\eqref{eq_Lcloudy}, \eqref{eq_extinctionCoef} and~\eqref{eq_albedo}, {\sc Mixclask} invokes {\sc Skirt} to recompute the mean intensity of the ISRF, $4\pi\lambda J_{\lambda}$, and the process is iterated until convergence as illustrated in Figure~\ref{Fig_flowchart}.

More precisely,
{\sc Mixclask} asks the user to provide a set of wavelengths or wavelength intervals, $\lambda_0$, $\lambda_1$, and it stores an average field\footnote{If a single wavelength is given, then $\langle \lambda J_{\lambda} \rangle (\lambda_0) = \lambda J_{\lambda}(\lambda_0)$ is just the mean intensity at that wavelength.}
\begin{equation}
    \langle \lambda J_{\lambda} \rangle(\lambda_0, \lambda_1)
    = \frac{1}{\lambda_1 -\lambda_0}\int_{\lambda_0}^{\lambda_1} \lambda J_\lambda\ \dd\lambda
    \label{eq_MagnitudeForConvergence}
\end{equation}
for each region and iteration.
The algorithm stops when the difference between percentiles $P_{16}$ and $P_{84}$ of the stored values (excluding the first iteration, where the ISM is neglected) is small compared to the median value $P_{50}$:
\begin{equation}
    \frac{P_{84} - P_{16}}{2\sqrt{N_i-1}} < \epsilon P_{50},
    \label{eq_convergence}
\end{equation}
where $N_i$ denotes the current iteration number, and $\epsilon$ is a tolerance parameter set by the user. 

By default, {\sc Mixclask} uses the interval $(\lambda_0, \lambda_1) = (10, 70)$\,nm to ensure convergence in the extreme ultraviolet (EUV), which is particularly interesting because of its physical relevance in the ionisation balance, and challenging in terms of convergence due to photon absorption by neutral hydrogen.
We use $\epsilon = 0.1$, the default value, in order to achieve a compromise between accuracy and computation time.
Expression~\eqref{eq_convergence} is tested for every spatial region, and {\sc Mixclask} performs another iteration of the algorithm until all of them have converged.
Note that, once convergence is reached, the percentiles $P_{16}$, $P_{50}$, and $P_{84}$ should approach an asymptotic value, and the relative difference with respect to the median is expected to decrease approximately as $\sim N_i^{-1/2}$.

The exact number of iterations that will be carried out depends on the selected tolerance $\epsilon$, the number $N_\gamma$ of photon packets used by {\sc Skirt} in each iteration, their probability distribution across the wavelength range, and the specific details of the problem at hand.
Hereinafter, this work uses $N_\gamma=10^7$ photon packets, and we enforce that approximately $50\%$ of them are uniformly distributed between $10$ to $90~$nm (see Appendix~\ref{sapp_skirt}).
As will be discussed below, condition~\eqref{eq_convergence} is satisfied in most regions after a few iterations, and the total run time is dominated by those zones that are more difficult to model.

In addition to testing for convergence, we also compute the percentiles at each individual wavelength and use them to provide an estimate of the statistical uncertainties in the final ISRF:
\begin{equation}
    \frac{\Delta \lambda J_\lambda(\lambda)}{\lambda J_\lambda(\lambda)} = \frac{P_{84}(\lambda) - P_{16}(\lambda)}{2P_{50}(\lambda)\sqrt{N_i-1}}.
    \label{eq_relativeError}
\end{equation}
As it will be shown below, these values are typically below the requested tolerance, intentionally based on a particularly problematic wavelength range.

\subsection{Test case: Spherical {\sc Hii} region}
\label{sec_testing}

\subsubsection{Outline of the simulations}

In order to test the scheme proposed in Figure~\ref{Fig_flowchart}, we study a case whose results may be compared against a standalone {\sc Cloudy} simulation.
For that purpose, we designed a test consisting on a blackbody at $3\times 10^4$\,K with a total luminosity of $5\times\,10^4 \,$L$_{\odot}$, surrounded by a spherical ISM shell between $10$ to $40$\,pc with a Hydrogen density of $1.0$\,cm$^{-3}$.
This test can be interpreted as a main-sequence B star inside a H{\sc ii} region. 

\begin{figure}
    \centering
    \includegraphics[width=0.4\textwidth]{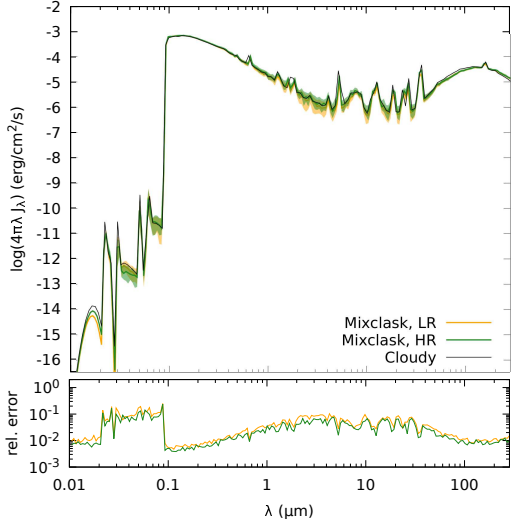}
    \caption{The upper panel shows the predicted spectrum of a blackbody after its radiation traverses a spherical shell from $10$ to $40$\,pc.
    The median ISRF returned by two {\sc Mixclask} runs with different spatial resolution is plotted as a solid coloured line, while a shaded area illustrates the $16$ and $84$ percentiles.
    The continuum transmitted by {\sc Cloudy} (solid black line) has been downsampled to have the same resolution as our results.
    The bottom panel displays the statistical error obtained from~\eqref{eq_relativeError}.}
    \label{Fig_BB_spectra}
\end{figure}

\begin{figure}
    \centering
    \includegraphics[width=0.45\textwidth]{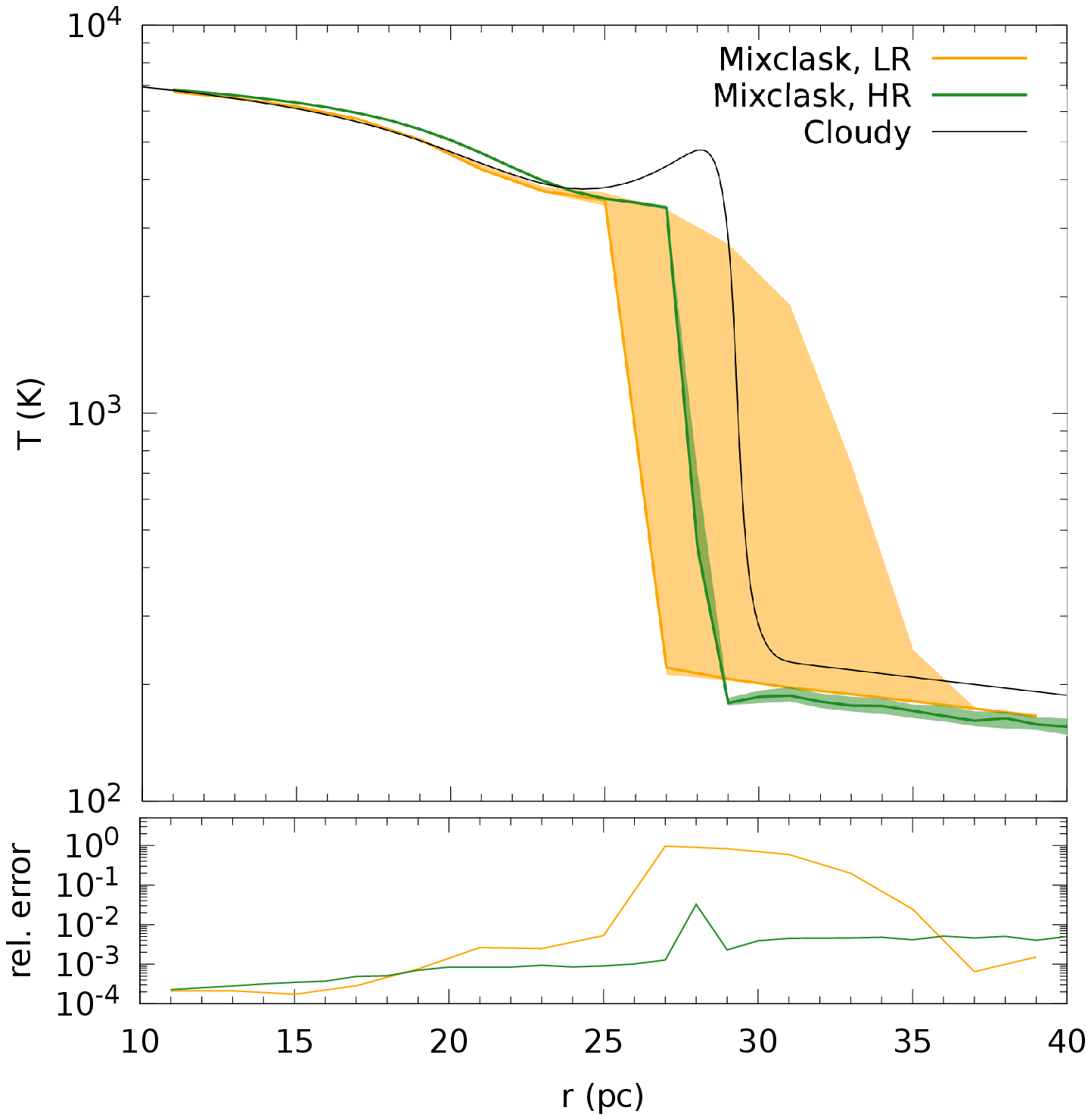}
    \includegraphics[width=0.45\textwidth]{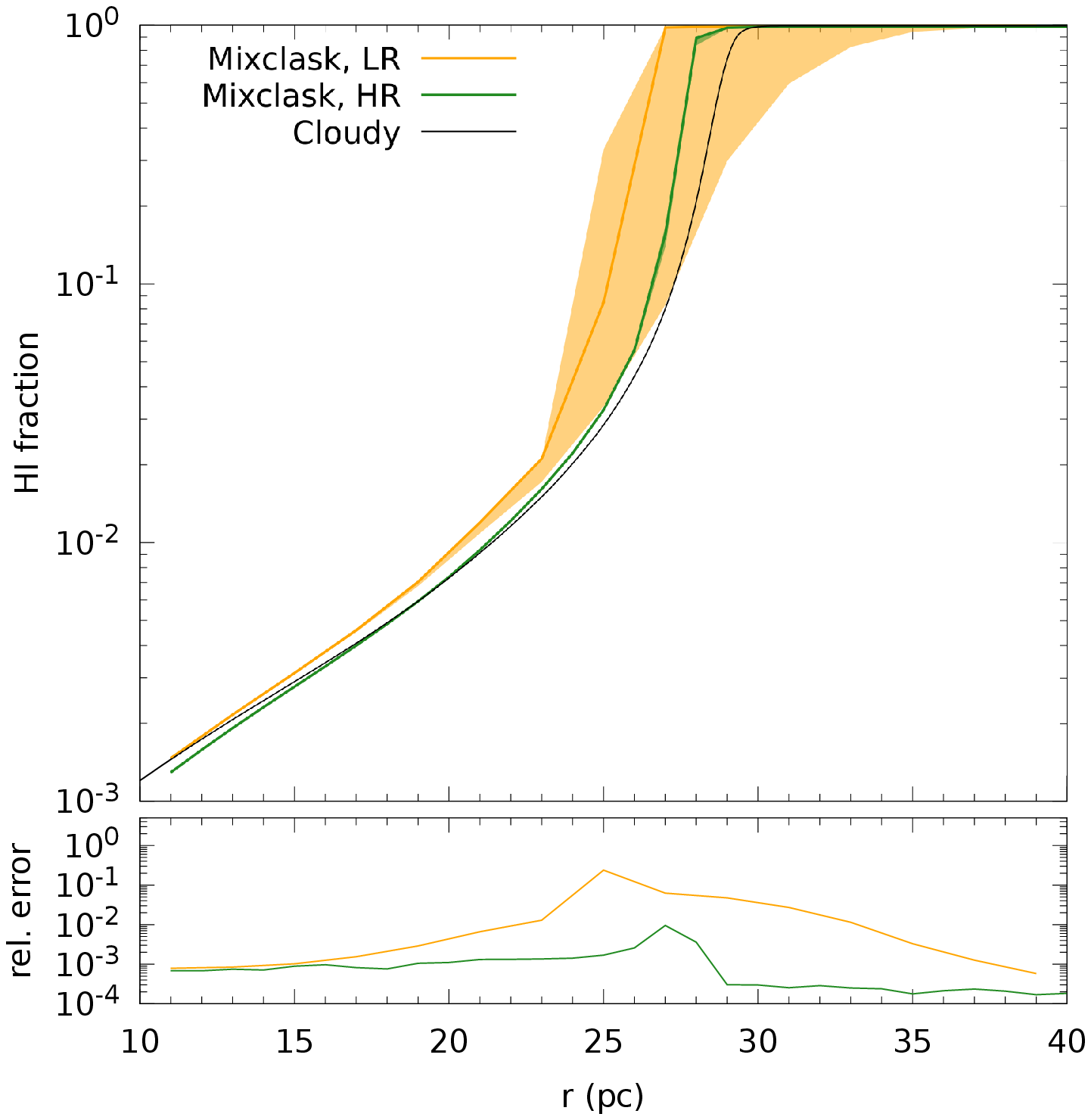}
    \caption{
    Predicted temperature and atomic hydrogen fraction profiles, and their relative errors~\eqref{eq_relativeError}, for a blackbody whose radiation traverses a spherical shell. }
    \label{Fig_BB_structure}
\end{figure}
 
\begin{figure}
    \centering
    \includegraphics[width=0.45\textwidth]{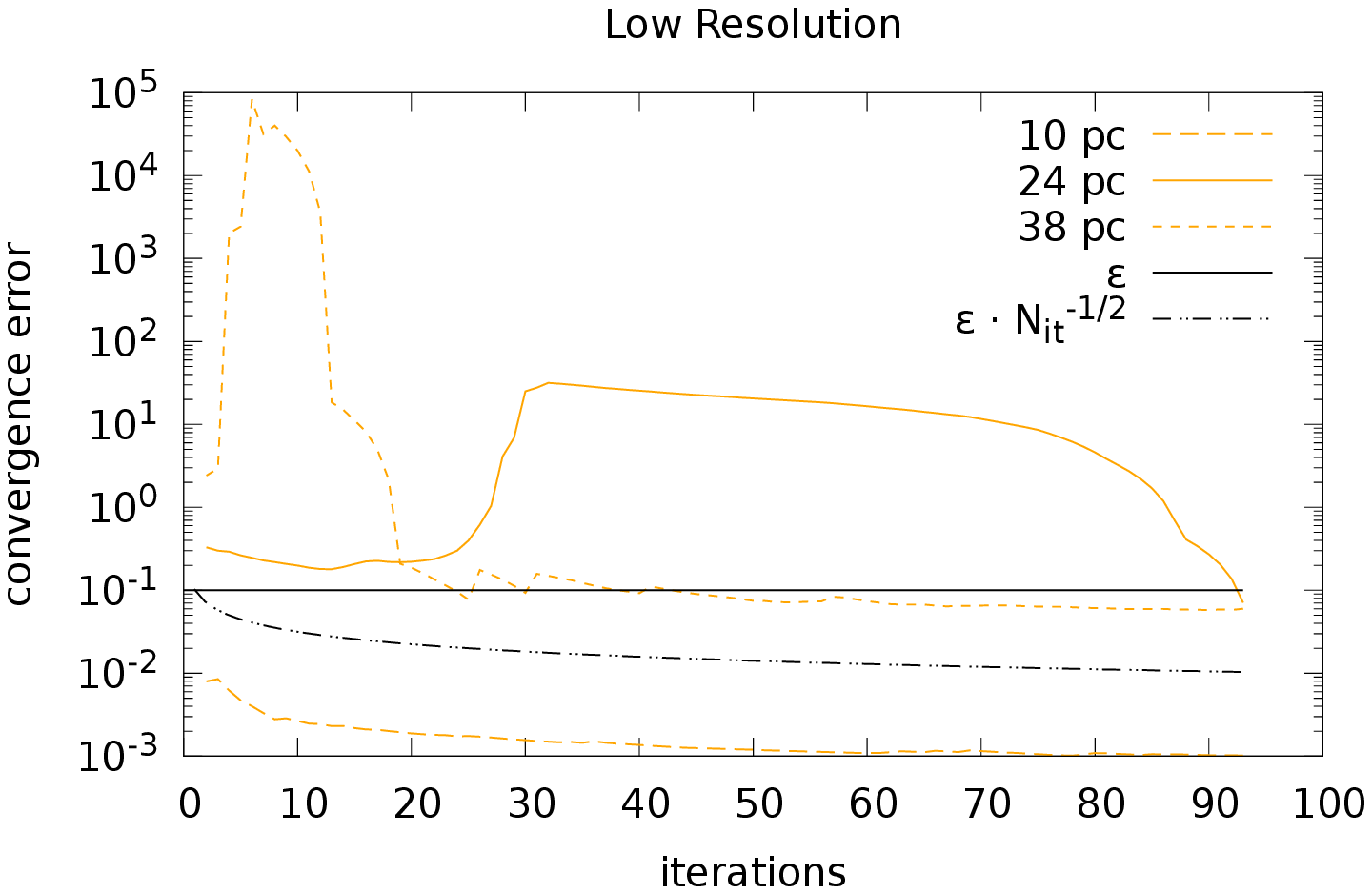}
    \includegraphics[width=0.45\textwidth]{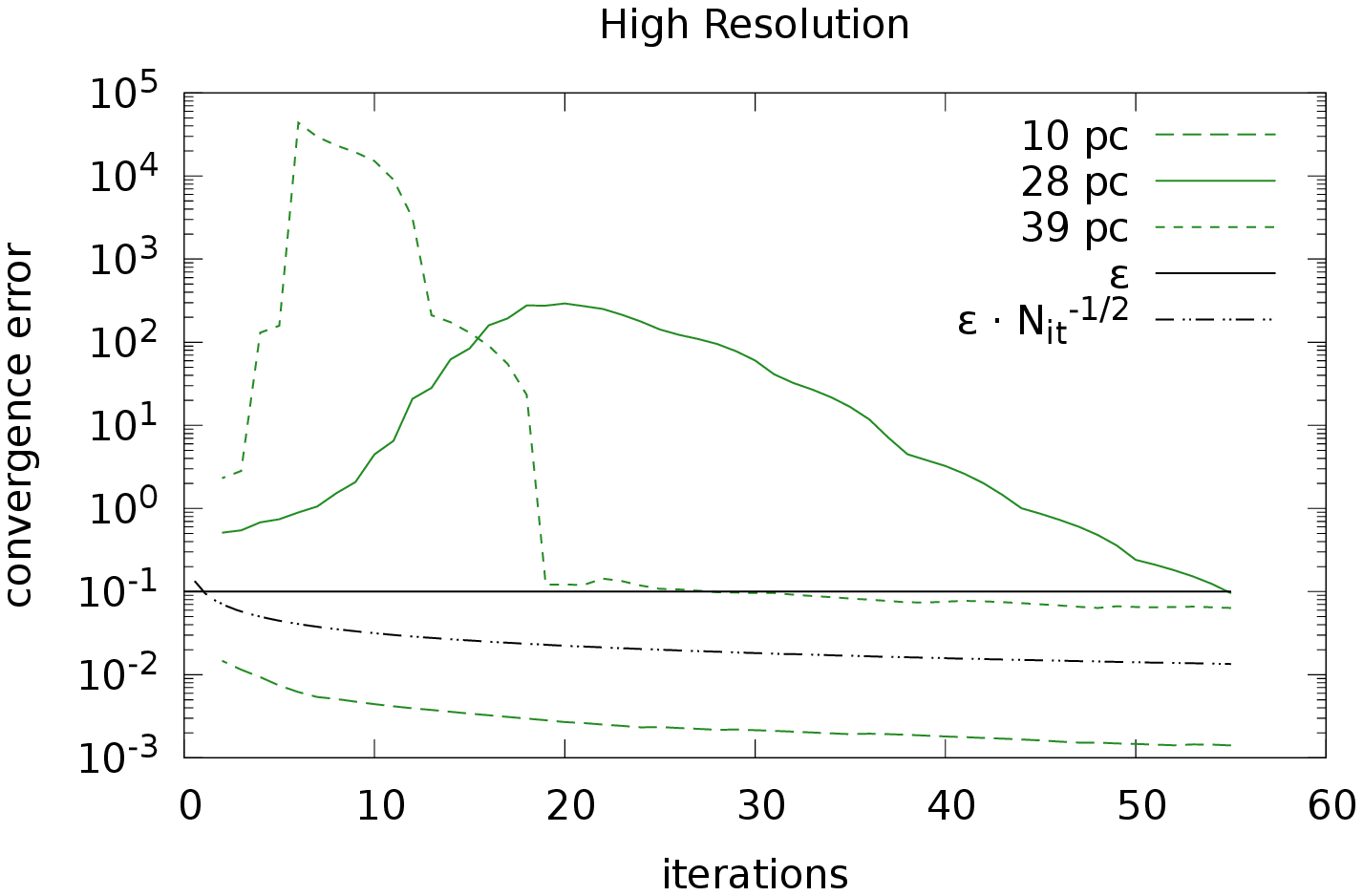}
    \caption{
    Convergence error, i.e. equation~\eqref{eq_convergence} divided by the median, as a function of iteration number, for different spherical regions (coloured lines).
    The requested tolerance, $\epsilon=0.1$, is shown as a black solid line, whereas the black dot-dashed line illustrates the expected $N_{it}^{-1/2}$ asymptotic behaviour.
    } 
    \label{Fig_convergence_BlackBody}
\end{figure}

The exact input file for the {\sc Cloudy} simulation is provided in appendix~\ref{app_cloudyControl}.
The code is set for a spherical cloud, with an initial radius of $10$\,pc, and it stops when the cloud thickness reaches $30$\,pc. Neither custom abundances nor dust are given as inputs, and therefore the default solar and ISM values in {\sc Cloudy} are used (see table~\ref{Table_inputs}).
We ask the code to iterate this simulation until convergence is reached.

In {\sc Mixclask}, the spherical shell is divided into several concentric regions whose width is $2$ (LR) and $1$\,pc (HR).
We would like to highlight that, as described in~\ref{sec_flowchart}, {\sc Mixclask} uses {\sc Skirt} to estimate the mean radiation field at the centre of each ISM region, and then it asks {\sc Cloudy} to solve plane-parallel slabs with the appropriate half-widths to compute their emission spectra and opacity in each iteration.
The spherical geometry of the regions is only considered by {\sc skirt}, when it performs the full radiative transfer to estimate the ISRF.
The wavelength ranges from $0.01$ to $300\,\mu$m in 200 logarithmic steps, and the default convergence criterion~\eqref{eq_convergence} is used.

\subsubsection{Mean intensity and photoionization structure}

The mean intensity field at the external boundary of the {\sc Hii} region (i.e. $40$\,pc from the radiating source) obtained from our code is compared in Figure~\ref{Fig_BB_spectra} with the transmitted continuum computed by the standalone {\sc Cloudy} simulation.
The spectrum shows three distinct regions: a heavily absorbed region, due to the presence of gas and dust, for energies below $13.6$\,eV ($\approx 0.091$\,$\mu$m); then the incident blackbody spectrum; and finally the dust (grains+PAH) emission in the infrared regime.
The agreement between the results of both codes is excellent throughout the whole wavelength range, regardless of spatial resolution.

Figure~\ref{Fig_BB_structure} displays information about the photoionization structure of the spherical ISM shell, in terms of the temperature and the {\sc Hi} fraction.
In contrast with mean intensities of Figure~\ref{Fig_BB_spectra}, increasing the number of {\sc Mixclask} regions has an impact on the predicted location of the ionization front, as well as the associated statistical uncertainties.

\subsubsection{Convergence analysis}

Figure~\ref{Fig_convergence_BlackBody} displays the behaviour of our convergence test statistic, equation~\eqref{eq_convergence} divided by the median, against the number of iterations until each simulation reaches convergence ($N_{it} = 93$ for the LR simulation and $N_{it} = 55$ for the HR case).
For each graph, we have selected three representative regions: the initial one, the precedent to the ionization front, and the outermost region.

On one hand, the very first region at $10$\,pc converges in the first few iterations, following a slope proportional to $N_{it}^{-1/2}$.
This result is expected, as the initial regions are so close to the ionizing radiation that it is barely absorbed.

On the other hand, the outermost regions achieve convergence between iteration $20$ and $30$ for both simulations, showing a sharp rise and decline before reaching the asymptotic regime.
In this case, ionizing radiation is heavily suppressed, and a certain number of iterations is needed to properly account for the effect of the preceding regions.
The ionization front is the region that dominates the number of iterations needed for convergence.
Radiation in that region is partially absorbed, and the transition between hot, ionised and cold, neutral gas (and therefore, the ISM optical properties) is very steep.

\section{The interstellar radiation field of the Milky Way Galaxy}
\label{sec_science}

Now that the validity of our scheme has been tested, let us address the feasibility of computing the ISRF of MWG from a self-consistent CEM.
In contrast to the usual approach of assuming a predefined analytical distribution for the stars and dust in the Galaxy, and fitting their parameters to the total intensity observed from the Earth at different wavelengths, we
start from a CEM to predict the relevant physical properties of stars and gas and their corresponding spatial distributions along the disk, and then {\sc Mixclask} is used to compute the ISRF.

With our technique, we improve the predictions of classical CEM, adding the radiation field to the results regarding the different stellar populations different phases and ISM phases.
We highlight the added value of our code with respect to {\sc Cloudy} and {\sc Skirt}, since it makes possible to consider all the information about the ISM (in particular, the chemical compositions of the gas and dust) provided by the CEM, as well as the stellar spectral energy distribution, as an input of the two-dimensional radiative transfer problem indicated in Figure~\ref{Fig_flowchart}.

We summarise in next section~\ref{ssec_CEM} the two CEM that we use to predict the distribution of gas and stars.
The {\sc Mixclask} setup for this calculation is described in~\ref{ssec_MixclaskSetup}, including the generation of the stellar spectra, as well as the composition of the ISM gas and dust phases.
We refer the reader to the references provided in these subsections for more details about each aspect.
Finally, the results are discussed in~\ref{ssec_results} against other works calculating the ISRF from the optical to the IR range of the spectrum~\citep[e.g.:][]{Mathis+83, Black87, PorterStrong06, Delahaye+10, Popescu+17}.

\subsection{Chemical evolution models}
\label{ssec_CEM}

CEM have been developed at early times \citep{Talbot_Arnett_1971, Talbot_Arnett_1973, Searle_Sargent_1972, Tinsley_1980, Pagel_1981, Pagel_1989} to study the evolution of Hydrogen, Helium and Metals across cosmic times from the Big Bang Nucleosynthesis (BBN) abundances to the present day one. 

Models solve an equation system to calculate the time evolution of different phases (stars and gas as basic ingredients) that compose the total mass or density in a galaxy or region within a galaxy, and the corresponding abundances of elements, which are created in the interior of stars and ejected to the interstellar medium when these stars die. 
The elements are incorporated to the subsequent generation of stars, at a different rate depending on the existing star formation rate, thus increasing the total metallicity or abundance of individual elements, in a given region or galaxy.
The first models tried to reproduce the G-dwarf metallicity distribution. Later, they aimed to predict radial gradients of elemental abundances of galactic discs as observed for C, O, N and Fe.

These elemental abundances are estimated through the measurements of emission lines from the gas, which is ionized by the hot young stars located in the well known {\sc Hii} regions. The precise measurements need the detection of the Oxygen line [O{\sc iii}]$\lambda$4363, with which the electronic temperature, $T_{e}$, of the region is determined. 
Once obtained, elemental abundances of O, N, S and others are easily obtained from the ratio between the corresponding emission line intensities. In other cases, the use of empirical calibrations give good estimates of these abundances \citep[see][for a review about the subject]{angel2010}.  
Other data useful to constraint chemical evolution models refer to the total, the stellar, the gas of atomic or molecular components, and the star formation rate surface densities, measured with different techniques, as dynamical methods for the first case, the luminosity in some bands in the second one, data in the radio band for obtaining the H{\sc i} mass with the 21\,cm line, or for measuring the CO line flux to calculate the mass of H$_{2}$. The star formation rate is usually obtained from the H$_\alpha$ flux, although there exist other calibrators, too. 

With all these observation data to fit, a large number of CEM have been developed in the last decades \citep[e.g.][]{Matteucci_Francois1989,Ferrini+1994,Molla+1996,Chiappini1997,Nieva_Przybilla_2012, Molla+2015, Hasselquist+_2017, Hasselquist+_2019, Queiroz+_2021, Hasselquist+_2021}, with increasing complexity in their ingredients, as infall from the intergalactic medium, outflows from negative feedback, different assumptions about the star formation law, stellar lifetimes, Initial Mass Function (IMF), stellar ejecta/yields and supernova rates.
In this work we use two recent chemical evolution models of a Milky Way-like galaxy (MWG), as detailed in next subsections.

\subsubsection{\sc Mulchem}

We have here used the results from the {\sc Mulchem} model \citep{Molla+2023} for the Galaxy disk. 
This model computed the chemical evolution for the MWG, assuming that Our Galaxy is divided in two zones, halo and disk, with a total dynamical mass of $9.55\times\,10^{11}\,M_{\sun}$. 
This mass is distributed in halo and disk regions following the \citet{salucci07} equations, which give the rotation velocity curves for both components as well as their radial distributions. 
The initial mass is assumed to be gas located in the halo, and to infall over the equatorial plane, forming a disk, with a rate inversely proportional to a collapse time scale $t_{\mathrm infall}(R)$. This last one is calculated in that way that the mass in each radial region of the disk at the present time be the observed value \citep[see details in][]{Molla+2016}.
The disk is divided into concentric rings at a radius $R$ with half height $h=0.2$\,kpc and half width $w= 0.5$\,kpc.

Star formation in the Galactic disk regions takes places in two steps, first forming molecular clouds from the diffuse gas, then forming stars by cloud-cloud collisions.
The efficiency of both processes is regulated by free parameters selected to reproduce the radial distribution of diffuse gas, molecular cloud, and star formation surface densities. 
Elements appear in the interstellar medium as a consequence of their ejection by stellar winds, planetary nebulae or supernova explosions. 
The stellar yields used are those from \citet{cristallo2011,cristallo2015} and \citet{limongi2018} for low and intermediate mass stars, and massive stars, respectively, and~\citet{iwamoto99} for supernovae Ia yields.

The equation system, related with the change of mass from a phase to other along time, and that the code solves, is given in \citet{Molla+17}. The basic inputs of the model are the selected stellar yield sets, the IMF that we use, \citet{Kroupa2001} in this case, the chosen virial mass, that define the infall rate, given above, and the efficiencies to form molecular gas and stars. As a result, the model gives the time evolution of the masses in form of ISM material (diffuse or molecular phases), stars (low and intermediate mass, massive ones and remnants), star formation rate, total mass in the disk zone, supernova --core collapse and SN-Ia types-- rates and elemental abundances for 14 species: H, $^{2}$H, $^{3}$He, $^{4}$He, $^{12}$C, $^{13}$C, N, O, Ne, Mg, Si, S, Ca and Fe. 
The {\sc Mulchem} model used is calibrated to reproduce the observed radial distributions of diffuse and molecular gas, stars and star formation rate for the disk of MWG, as well as the age-metallicity relation and star formation history data of the Solar vicinity.
The evolution of C, N and O abundances in this region, and their radial gradients in the disk are also well fitted.
The data used for this calibration are summarized in \citet{Molla+2015}, and model details can be found in \citet{Molla+17,Molla+2023}.

\subsubsection{\citet{Millan-Irigoyen+20}}
\label{sec_MMA}

We have also used the models of \citet[][hereinafter MMA]{Millan-Irigoyen+20}.
Apart from considering the different gas phases that exist in the ISM (atomic, ionized and molecular), these galactic chemical evolution models follow the formation, growth and destruction of cosmic dust grains. 
Dust is a critical ingredient in order to correctly model the ISRF, as it is the main contributor to the absorption of starlight at optical and ultraviolet wavelengths \citep[to the extent that it may be considered a good tracer of the diffuse gas; e.g.][]{Eales+12,Groves+15}, and it dominates the emission in the infrared range~\citep[e.g.][]{Freudenreich98,Robitaille+12,Natale+21}.

The MMA models feature three free parameters that are specifically related to the dust physics:
the sticking coefficient $S$ (i.e. the probability that an atom gets stuck to a grain when they collide), the destruction timescale $t_{\rm dest}$ of dust grains due to different processes, and the characteristic particle number density $\rm n_{cloud}$ of Giant Molecular Clouds.
The value of these three variable parameters is assumed to stay constant throughout the galaxy, and different regions are fully specified by the total mass per unit area $\Sigma_{\mathrm total}$ that has been accreted over the cosmic history and the characteristic infall timescale $t_{\mathrm infall}$.
Thus, $\rm \Sigma_{total}$ determines the sum of the stellar and ISM surface densities in each region at the present time, and ${\rm t_{infall}}$ is the parameter that regulates the evolutionary state, with shorter times yielding more chemically evolved \citep[gas-poor, metal-rich; see e.g.][]{Ascasibar+15} regions.

For this specific test case, we have set the free parameters that regulate the dust physics to $S=0.7$, $t_{\mathrm dest}=1.3$~Gyr and $\rm n_{cloud} = 5000$ cm$^{-3}$, and we have run a series of chemical evolution models to represent the same disk regions in terms of $\{R, w, h\}$ as {\sc Mulchem}.
For the radial variation of the total surface density $\Sigma_{\mathrm total}$ and infall timescale $t_{\mathrm infall}$, we have chosen the simple parameterisation:

\begin{eqnarray}
    \Sigma_{total} &=& (1000 ~\text{M}_{\odot} \text{pc}^{-2})~e^{-R/3.3~\text{kpc}} \\
    t_{infall} &=& (2~\text{Gyr}) ~(\frac{R}{8.5~\text{kpc}})^{\frac{1}{6}}
\end{eqnarray}
that has been manually optimised to reproduce the observed radial distributions of gas and stars given in \citet{Molla+2015} for the MWG.

Regarding other parameters, the MMA models also adopt the \citet{Kroupa2001} IMF.
The stellar yields used are taken from MOL15, selecting \citet{gavilan2005,gavilan2006} for low and intermediate mass stars, \citet{ww95} for massive stars, and \citet{iwamoto99} for supernovae Ia.
For the dust, yields are taken from \citet{ventura2012a, ventura2012b, criscienzo2013} and \citet{dellagli2017} for AGB stars, from \citet{marassi2015} for massive stars dying as core-collapse supernovae, and from \citet{nozawa} for the Supernoave Ia.
The reader is referred to \citet{Millan-Irigoyen+20} for further details.

\subsection{{\sc Mixclask} setup}
\label{ssec_MixclaskSetup}

The geometry of the galaxy is modelled as a sum of concentric rings whose half widths $w$ and heights $h$ are consistent with the choice adopted in the chemical evolution models.
Population synthesis models must then be used to estimate the luminosity emitted by the stars in each ring from the age and metallicity histories returned by the CEM. The ISM mass and Hydrogen number density are derived from the predicted surface densities at the present time, and their chemical composition and dust content are also provided by the CEM output.

\subsubsection{Stellar population synthesis}
\label{ssec_SPS}

From the star formation rate (SFR) given by each CEM at every radial region, one can compute the stellar spectra by using the well known technique of evolutionary synthesis models, where the specific luminosity of each region, as a function of wavelength, is obtained by adding the contribution $L_{\rm SSP,\lambda}$ of Simple StellarPulations (SSP) with different ages and metallicities:
\begin{equation}
    L_\lambda(t) = \int_0^t \psi(t-t')\ L_{\rm SSP, \lambda}(Z(t-t'), t')\ \dd t'
    \label{eq_Stellar_L}
\end{equation}
where $\psi(t-t')$ and $Z(t-t')$ denote, respectively, the instantaneous star formation rate and gas metallicity at time $t-t'$, as provided by the CEM. The spectral energy distribution (SED) $L_{\rm SSP,\lambda}(Z, t')$ of a SSP of metallicity $Z$ and age $t'$ is computed by an evolutionary synthesis model from the individual luminosities of stars in different regions of the Hertzsprung-Russell (HR) diagram.
Here we have used the {\sc PopStar} evolutionary synthesis models\footnote{\url{https://www.fractal-es.com/PopStar/}} with a \citet{Kroupa2001} IMF. 
In particular, we have used the low resolution models from \citet{PopStar2009} instead of the newer high-resolution ones from \citet{millan-irigoyen2021}, since the former cover a broader wavelength range, from 9.1 to 1.6$\times\,10^{5}$\,nm. 
The resulting SED are given in solar luminosities by wavelength unit and by solar mass units (L$_\odot$\,\AA$^{-1}$\,M$_\odot^{-1}$).

\subsubsection{ISM composition}
\label{ssec_ISMcomposition}

Both CEM differ in some physical magnitudes and outputs.
On the one hand, {\sc Mulchem} gives particular ISM abundances $\Tilde{X}_i = M_i/M$ within the galaxy, regardless of whether these elements are forming grains or dispersed in the gas phase.
Therefore, we correct {\sc Mulchem} abundances to take into account the quantity depleted into grains and obtain the necessary mass fraction (in terms of the gas mass) used by {\sc Mixclask} and {\sc Cloudy}.
For each element $i$, the conversion into $X_i$ is:
\begin{equation}
    X_i = \frac{\mathcal{D}_i \Tilde{X}_i}{ \sum_{j} \mathcal{D}_j \Tilde{X}_j },
    \label{eq_MulchemMixclaskConversion}
\end{equation}
where $\mathcal{D}$ corresponds to the depletion factors given in the Table 7.8 of {\sc Cloudy} documentation (i.e. Hazy I), and the summation in the denominator, equivalent to the gas mass, runs only over the elements followed by {\sc Mulchem}.
We also take advantage of the depletion factors to compute the dust-to-gas ratio of each region:
\begin{equation}
    DTG = (\sum_{j} \mathcal{D}_j \Tilde{X}_j)^{-1} - 1
\end{equation}
On the other hand, the MMA model follows the global gas metallicity $Z$ and the mass ratio between \emph{grains} and gas, not including PAH.
We set the Helium mass fraction to $Y=0.24$ for all regions and convert the 'grain-to-gas' ratio to a DTG dividing it by $1-q_{\rm PAH}$.

Neither of the two models follows the PAH's {or cosmic rays.}
{For the former,} we assume $q_{\rm PAH} = 7\,\%$ for all regions.
This value is below the upper limit reported in~\citet{Galliano+18} for nearby galaxies, and it is motivated by the results of~\citet{Robitaille+12}, who found that this number, rather than the fiducial one $q_{\rm PAH} = 4.6\%$ used in the dust models from \citet{DraineLi07}, was required in order to provide a good fit to MWG observations.
For the latter, we use the default cosmic ray ionization rate given by {\sc Cloudy} for all regions, based on the H$^0$ ionization rates of~\citet{GlassgoldLanger74} and~\citet{Indriolo+07}.

\begin{figure}
    \centering
    \includegraphics[width=0.45\textwidth]{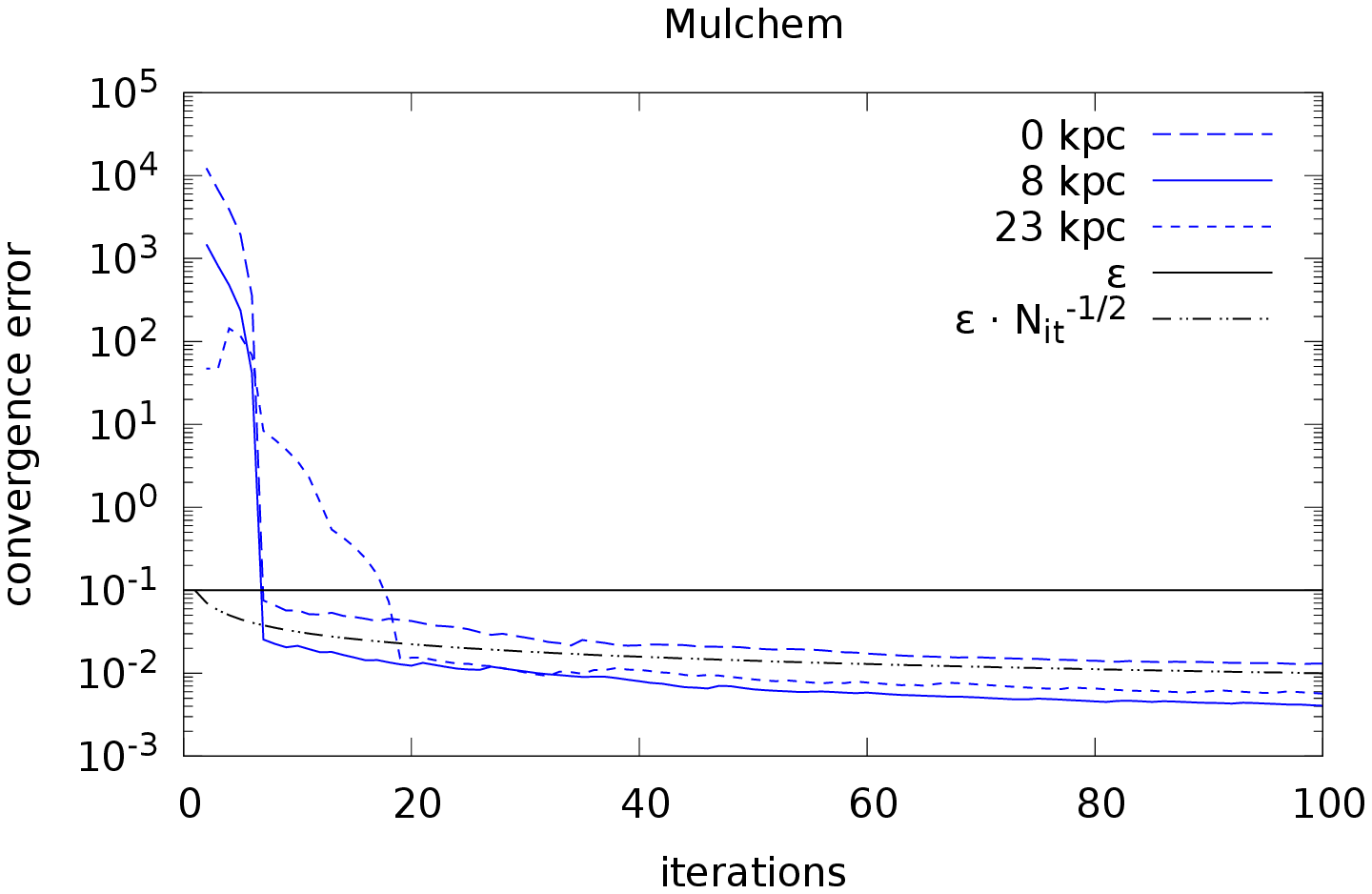}
    \includegraphics[width=0.45\textwidth]{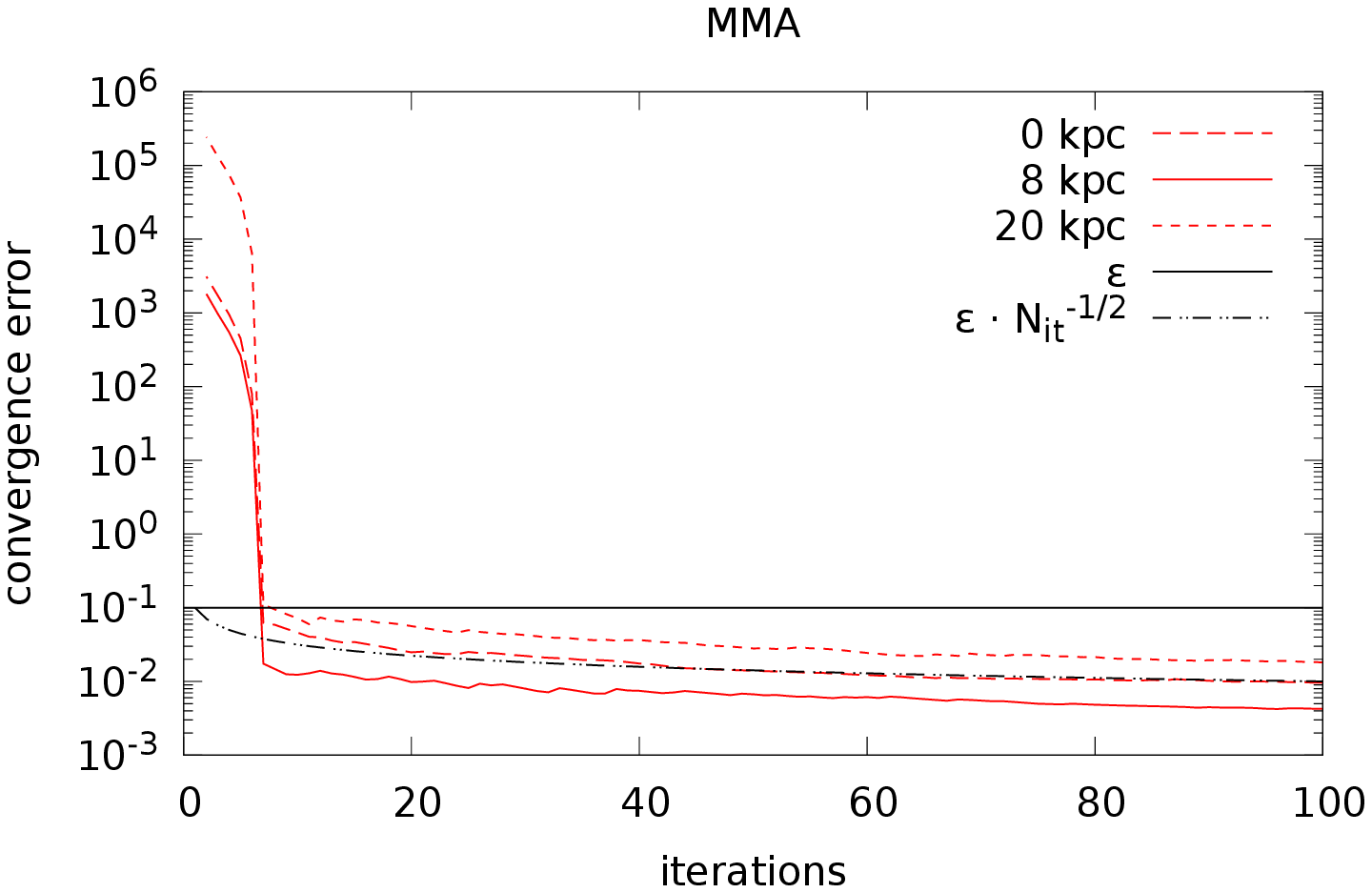}
    \caption{
    Values obtained from convergence per iteration against the number of iterations.
    coloured lines are equation~\eqref{eq_convergence} divided by the median, and they cross the $0.1$ black solid line (which is the tolerance imposed in~\ref{sec_testing}) before showing a curve proportional to $N_{it}^{-1/2}$ (black dot-dashed line).
    selected tolerance ($0.1$, black solid line), and that value divided by the square root of the number of iterations (black dot-dashed line).
    }
    \label{Fig_convergence_MWG}
\end{figure}

\begin{figure*}
    \centering
    \includegraphics[width=1\textwidth,height=0.41\textwidth]{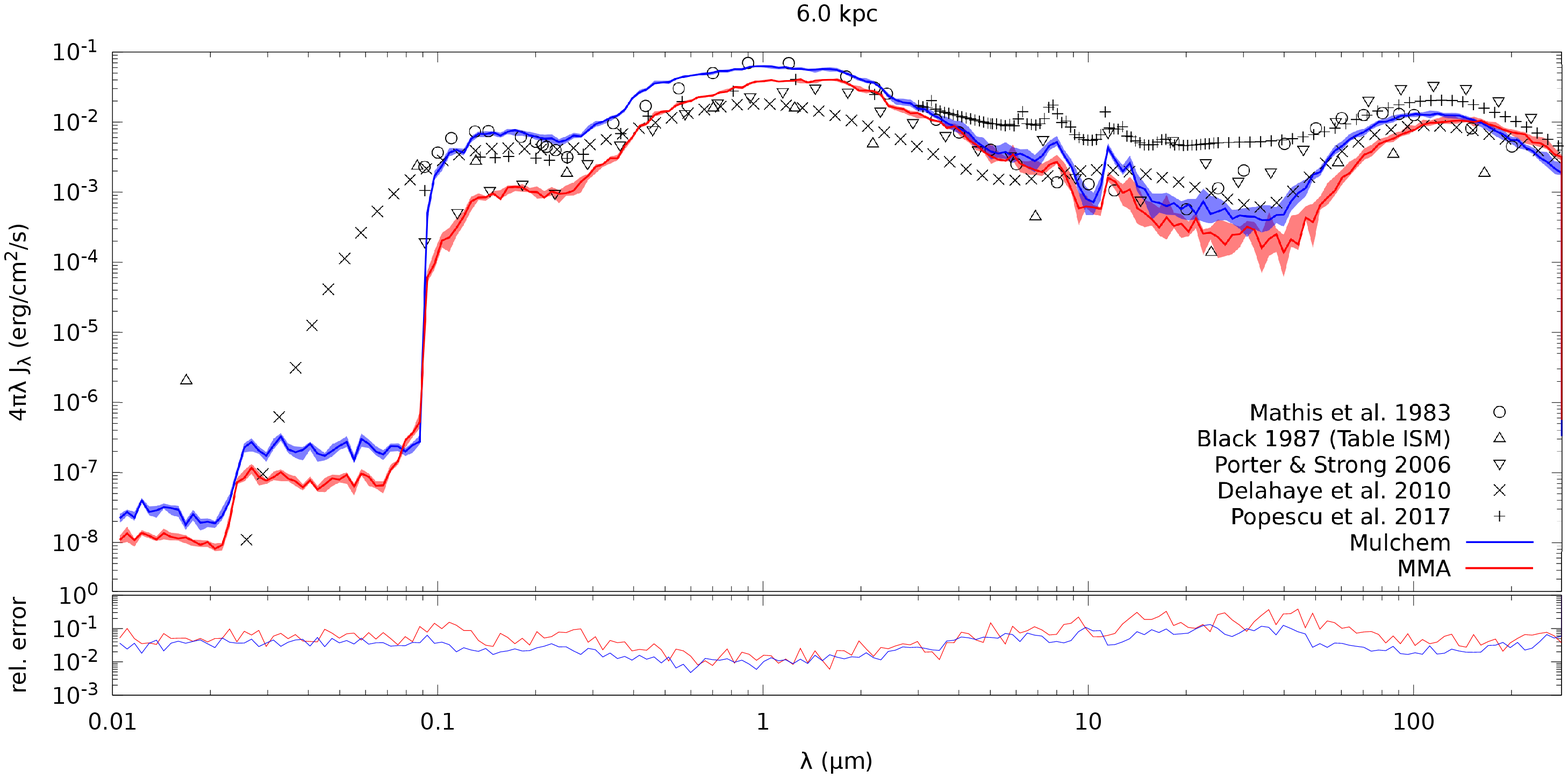}
    \includegraphics[width=1\textwidth,height=0.41\textwidth]{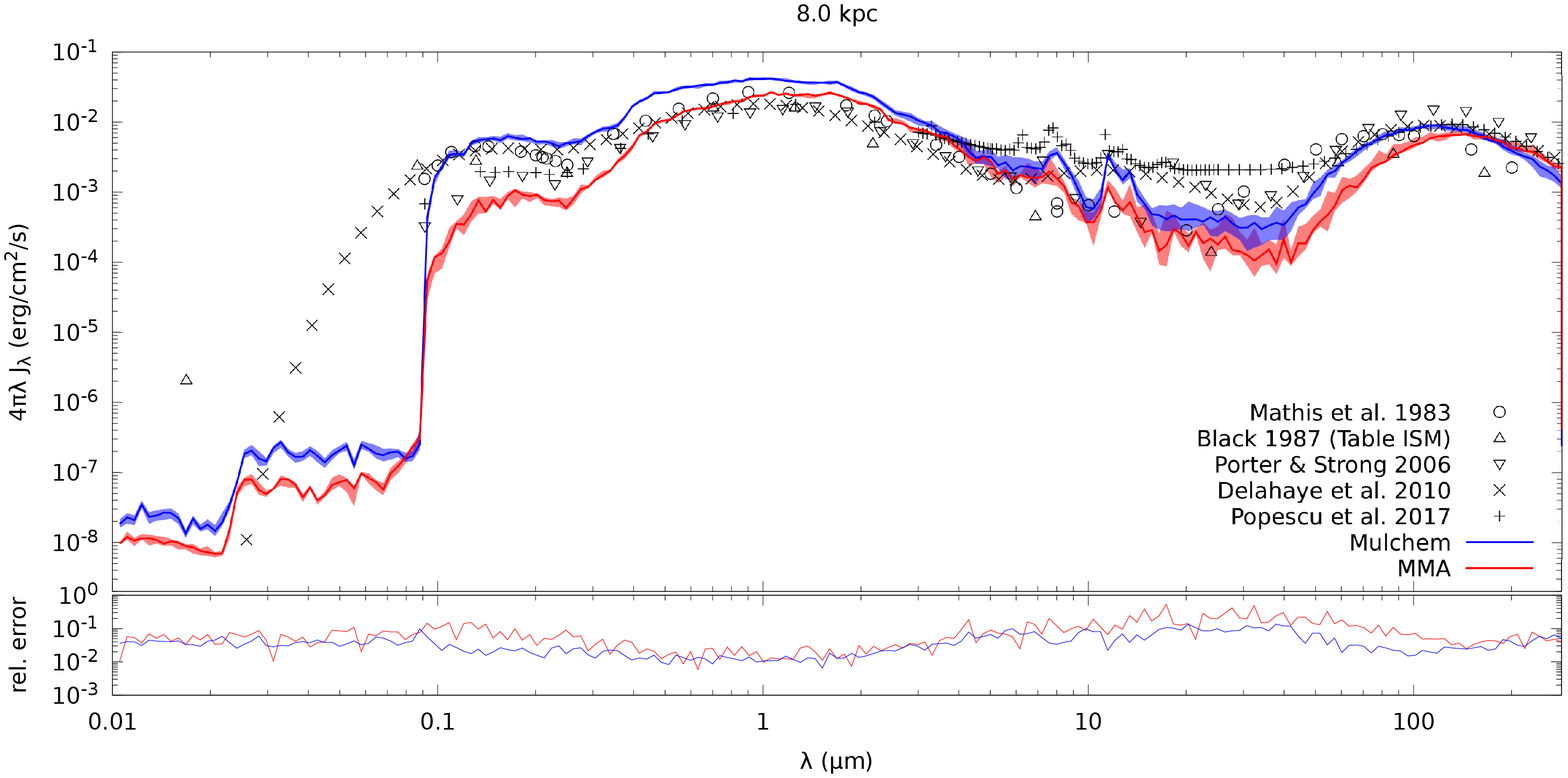}
    \includegraphics[width=1\textwidth,height=0.41\textwidth]{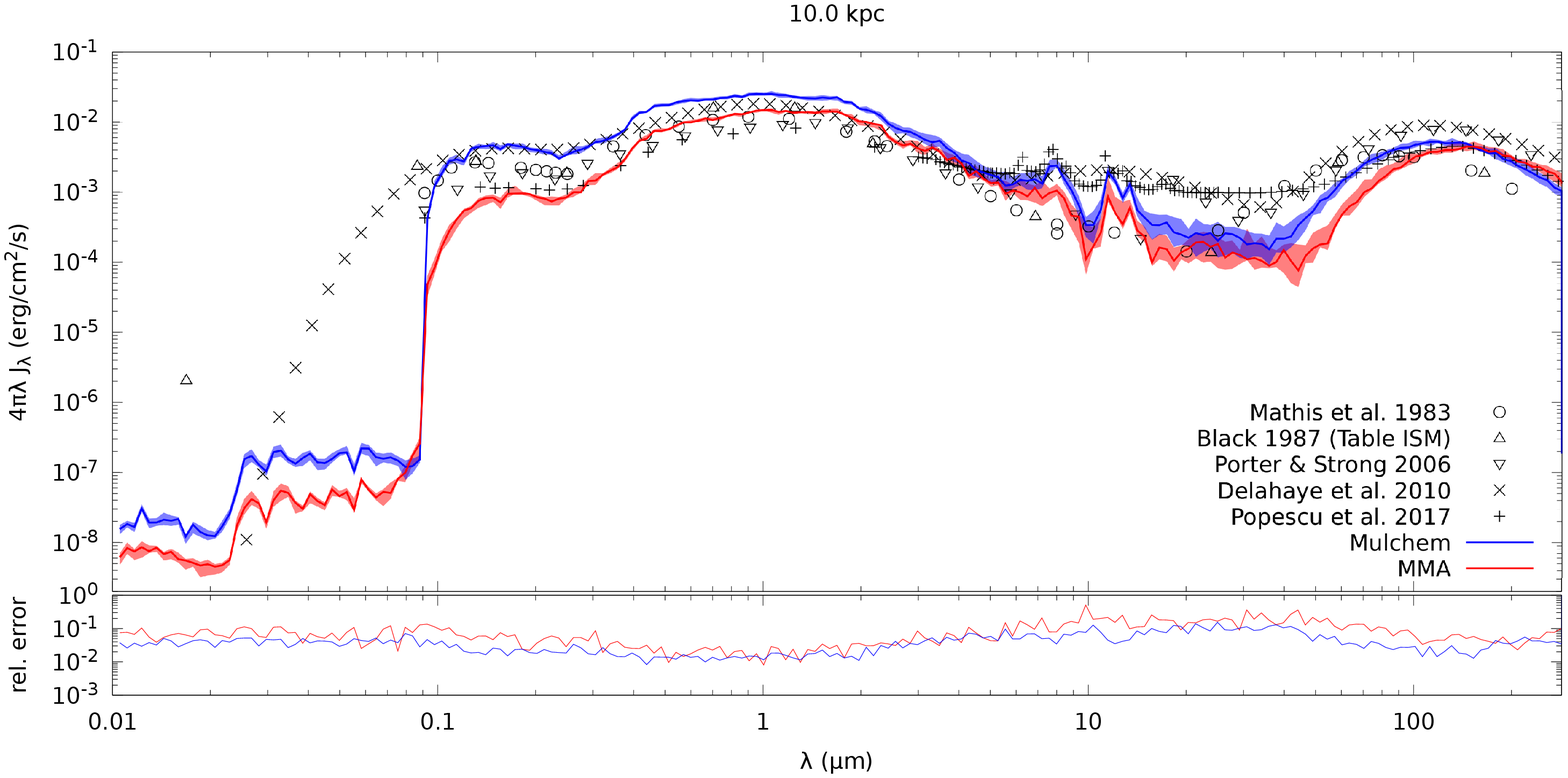}
    \caption{Interstellar radiation field at midplane for different locations near the Solar Neighborhood.
    Median ISRF predicted by {\sc Mixclask} from {\sc Mulchem} (blue line) and MMA (red line) chemical evolution models, with 84th and 16th percentiles indicated by shaded areas, compared to previous works in the literature (black data points).
    Lower insets show the relative errors following~\eqref{eq_relativeError}.}
    \label{Fig_solar_spectra}
\end{figure*}

\subsubsection{Convergence}

In order to study convergence, we first run both simulations with the default values described in~\ref{ssec_ConvergenceAndErrors}, $\epsilon=0.1$ and a wavelength range between $10$ to $70$\,nm.
We found that {\sc Mulchem} and MMA ISRF converged in $18$ and $7$ simulations, respectively, but we keep iterating until $N_{it} = 100$ to verify that the expected asymptotic behaviour, $\sim N_{it}^{-1/2}$, is reached afterwards.

In the same manner as Figure~\ref{Fig_convergence_BlackBody}, Figure~\ref{Fig_convergence_MWG} displays equation~\eqref{eq_convergence} divided by the median for both models.
This time, we select the Galactic centre, the solar neighborhood (i.e.: $8$\,kpc) and the outermost region ($23$\,kpc in  {\sc Mulchem} and $20$\,kpc in MMA) for plotting purposes.
The outermost region always dominates the number of iterations required to achieve convergence for $\epsilon = 0.1$.
If a lower tolerance was set, the Galactic centre would impose the most restrictive constraint in the case of {\sc Mulchem}.

From the results shown in Figure~\ref{Fig_convergence_MWG}, we conclude that the ISRF did indeed converge for $\epsilon = 0.1$.
Of course, a lower tolerance may be specified to obtain a better accuracy, albeit the number of iterations needed for reaching a much lower threshold may be impractical.
Moreover, the differences we find between the predictions of our models (see below) highlight that statistical errors are negligible in comparison to systematic uncertainties.

\subsection{Results and discussion}
\label{ssec_results}

\begin{figure*}
    \centering
    \includegraphics[width=0.49\textwidth]{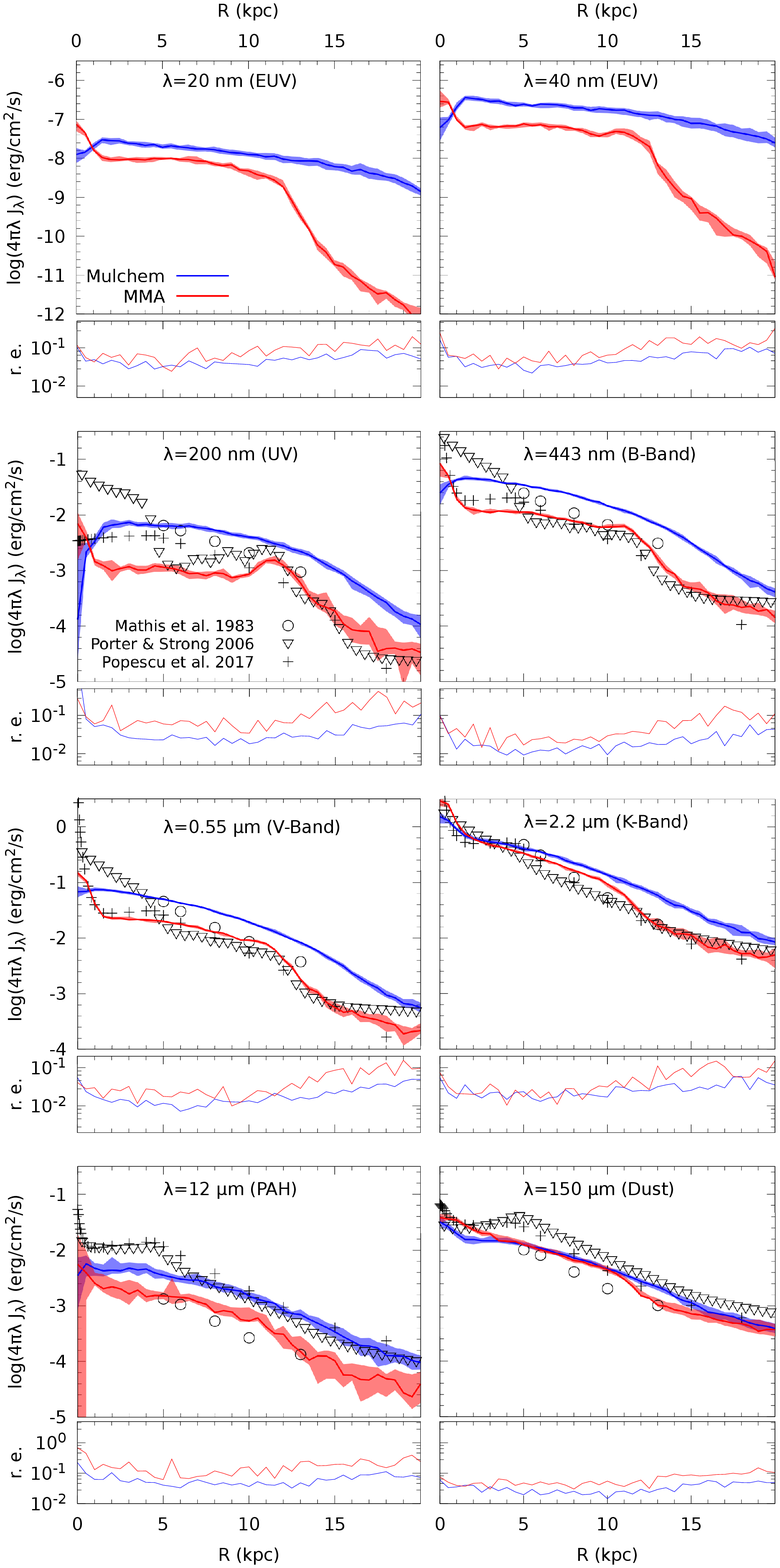}
    \includegraphics[width=0.49\textwidth]{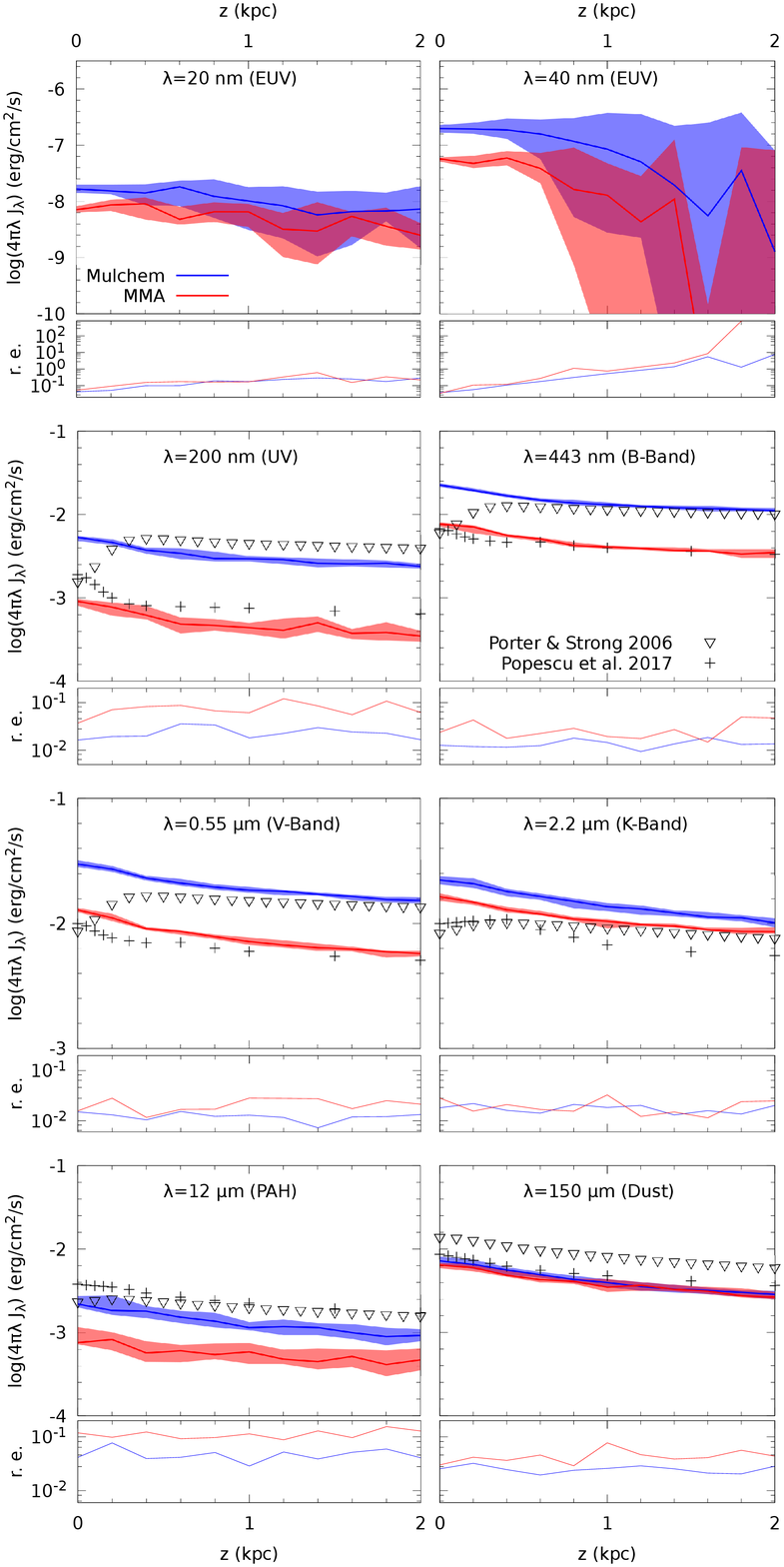}
    \caption{The two columns on the left: Radial profiles at midplane of the interstellar radiation field for a set of characteristic wavelengths. 
    The right columns are the vertical profiles of the interstellar radiation field at $R = 8$\,kpc for the same wavelengths.
    Median ISRF generated by {\sc Mixclask} from {\sc Mulchem} and MMA results are coloured as blue and red lines, respectively, whereas the shaded area represent the 16th and 84th percentiles. Lower insets show the relative errors following Eq.~\eqref{eq_relativeError}.
    }
    \label{Fig_profiles_spectra}
\end{figure*}

\subsubsection{Comparison against previous works}

We show in Fig.~\ref{Fig_solar_spectra} the mean intensity field predicted by {\sc Mixclask} at three regions located at the mid-plane of the disk and $R = \{6, 8, 10\}$\,kpc way from the galactic centre (top, middle and bottom panels, respectively).
We also add, for the sake of comparison, a representative sample of models of the Milky Way ISRF that cover the whole electromagnetic spectrum from $10$\,nm to IR wavelengths.
Reported values of the radiation energy density $\lambda u_\lambda$ are converted to average intensities by means of the usual relation $4\pi\lambda J_\lambda = c \lambda u_\lambda$, where $c$ is the speed of light.
As in Section~\ref{sec_testing}, {\sc Mixclask} results plotted in Figures~\ref{Fig_solar_spectra} and~\ref{Fig_profiles_spectra} are the median $4\pi \lambda J_{\lambda}$ (solid lines), and coloured areas correspond to percentiles $16$ and $84$.

The radial dependence of the mean intensity at the mid-plane is shown in the two left columns of Fig.~\ref{Fig_profiles_spectra} for eight different wavelengths, as labelled in each panel, between $0.02$ and $150$\,$\mu$m that are representative of the different constituents of the ISRF:
a) $0.02$ and $0.04$\,$\mu$m represent the Extreme Ultraviolet (EUV), where radiation from young stars is heavily absorbed by the interstellar gas;
b) $0.2$\,$\mu$m is the (less) absorbed ultraviolet light from the young stellar population; c) Johnson B, V and K bands for older stars; d) and $12$ and $150$\,$\mu$m in the infrared, where the emission PAH and grains peak, respectively, according to Figure~\ref{Fig_solar_spectra}.
The corresponding vertical profiles at $R = 8$\,kpc are plotted on the two right columns of this same Fig.~\ref{Fig_profiles_spectra}.

The models by \citet{Mathis+83,Black87} and~\citet{Delahaye+10} were initially proposed only for the Solar Vicinity \citep[or nearby, in case of][]{Mathis+83}, not applied to the whole Galaxy, and therefore we have opted for not representing \citet{Black87} nor~\citet{Delahaye+10} in Fig.~\ref{Fig_profiles_spectra}, while only the radial profile is presented for~\citet{Mathis+83}.
Nevertheless, we note that both of them have often been applied to other galactocentric distances.
\citet{Black87} is widely used in a variety of Galactic and extragalactic settings, as it is the default ISM radiation field adopted in {\sc Cloudy} (the user is able to specify an optional scaling parameter to change its overall normalisation, but the spectral shape is fixed), while the ISRF model by \citet{Delahaye+10} has received considerable attention in the astroparticle physics literature.
It is included, for instance, as one of the available options in the cosmic ray propagation code {\sc Dragon}~\citep{DRAGON2}, albeit most studies based on this code use more elaborate prescriptions, such as e.g. \citet[][]{PorterStrong06}, that provide information not only about the radial variation of the ISRF but also its vertical structure (Fig.~\ref{Fig_profiles_spectra}).
We also include the recent ISRF model by~\citet{Popescu+17}, where different axisymmetric models are assumed for the stellar emissivity and dust opacity.

We want to highlight that once again that there are two major differences between previous ISRF models in the literature and the new predictions presented in this work.
First, we use the results obtained from a physically-motivated CEM as an input for {\sc Mixclask} instead of calibrating a phenomenological model directly from observations.
Second, our detailed account of gas absorption enables the \emph{prediction} of the ISRF in the EUV range, in contrast previous studies of the Milky Way ISRF.

\subsubsection{Differences between chemical evolution models}

The predictions of the two CEMs considered in this work are broadly consistent with the trends obtained by other models, with our results within a range from about a factor of $2$ to an order of magnitude in the Solar neighbourhood, with the largest differences occurring in the mid-IR regime dominated by PAH emission.
They also feature different radial and vertical gradients for the ultraviolet and optical bands.
The fact that we do not observe an obvious systematic departure from previous results in any region of the electromagnetic spectrum (except perhaps the underestimation of the mid-IR emission discussed below) further validates the proposed approach.
Note that, at variance with previous reconstructions of the Galactic ISRF, our method is not fine-tuned to reproduce the observed emission, but to provide a physically meaningful and self-consistent picture of a MWG disk galaxy, describing not only its current state but its entire assembly history.

We do find significant differences between both chemical evolution models, which can be traced to the underlying physics. Thus, there is a discrepancy for high latitudes between the two models that amounts to a factor of $3-55$. Moreover, the precise shape of the ISRF at the Galactic centre is remarkably uncertain, with differences that can be as high as one order of magnitude in the UV between both models.
Due to the assumed mass distribution, based on the \citet{salucci07} rotation curve, {\sc Mulchem} tends to predict a higher stellar surface density than MMA, built upon the simple parameterization described in Section~\ref{sec_MMA}.
On the other hand, the conversion of gas into stars is less efficient in MMA, and therefore this model yields a larger amount of gas and dust at the present time, and thus a more absorbed contribution of young blue stars in the short-wavelength band of the spectra.
This feature is also reflected into the IR regime, where {\sc Mulchem} predicts systematically higher dust emission than MMA due to its higher stellar mass density, albeit the FIR-to-UV ratio, indicative of dust re-processing, is higher in MMA.

Regarding the spatial distribution of the ISRF, MMA predicts a flatter radial gradient compared with {\sc Mulchem} up to $R \sim 10$\,kpc.
At that point, the profile displays a sharp truncation due to the different star formation prescription, that produces an inefficient star formation there, and that features a fairly sharp transition when a critical metallicity is reached \citep{Millan-Irigoyen+20}.
The vertical structure is fairly similar in both models, but one must bear in mind that this is hardly surprising, as the same geometry has been imposed in {\sc Mixclask} to their concentric rings.
In terms of the normalisation, the discrepancy between {\sc Mulchem} and MMA is quite similar to the systematic offset observed between \citet{PorterStrong06} and \citet{Popescu+17}.
Differences in the stellar yields may also play a role~\citep{Molla+2015, Cote+16} and lead to subtle effects in the chemical evolution of the galaxy that ultimately reflect on the predicted ISRF.

\subsubsection{Limitations}

One of the most relevant differences between our two models is the treatment of dust extinction, which has a significant influence at UV and IR wavelengths.
Unfortunately, none of them incorporates a prescription to directly follow the evolution of PAH, and this is why we have selected the constant value $q_{\rm PAH} = 7\,\%$ as a crude, first-order approximation, following~\citet{Robitaille+12}.
The physical properties of dust and PAH are an open problem and an active field of research, and our results suggest that such a simple prescription fails to reproduce the observed radiation field in the mid-IR range.
The solution is at present far from obvious, since it does not seem likely that the abundance of PAH can be increased much further than the adopted value, and we did not detect any significant issues with the dust temperatures and emissivities estimated by {\sc cloudy}.
It is possible that the discrepancy is related to the abundance and properties of very small grains (size between 2 and 20~nm), but a much more thorough analysis would be required in order to address this question. 

One may thus conclude that the proposed tool has been successfully applied to \emph{predict} the intensity of the ISRF from a self-consistent chemical evolution model, albeit more physical ingredients should be considered in order to accurately reproduce all the available observations and understand the underlying processes at stake.
For instance, the lack of adequate stellar templates prevents us from modelling the ISRF in the X-ray regime, which may be important in regulating the gas properties in optically thick regions of the ISM.
Other heating sources, such as shock waves and strong turbulence, should also be taken into account.
Cosmic rays and magnetic fields are also known to play a pivotal role, both in terms of correctly predicting the gas density and temperature, as well as the emissivity in the radio and gamma-ray bands.
The axial symmetry assumed in both our CEM is also an important limitation, and a more detailed treatment of the three-dimensional structure of the ISM would be highly desirable, in particular should shocks or turbulence are included.

Bearing all these caveats in mind, we would nevertheless like to argue that the approach presented here has significant advantages over a purely phenomenological fit (e.g. based on analytical functions), as it paves the way towards a more physics-oriented discussion.
All the components and processes that co-exist in a galaxy are deeply connected, and a fully self-consistent model, even if it is difficult to achieve, is better constrained and offers much more information than a description based on independent, freely varying parameters.

\section{Conclusions}
\label{sec_Conclusions}

This work predicts the interstellar radiation field (ISRF), from $10~$nm to $300~\mu$m, from the output of a chemical evolution model (CEM).
To achieve this goal, we develop an iterative scheme, implemented in a publicly available code dubbed {\sc Mixclask}, which combines {\sc Cloudy}\citep{Cloudy17} and {\sc Skirt}\citep{Skirt9}.

Based on the geometry defined by the CEM, the simulation domain is divided into regions that contain stars and ISM material.
Once the spectral energy distributions of the stellar populations and the chemical composition of the gas and dust that comprise the ISM are specified, {\sc Mixclask} returns the mean intensity $4\pi\lambda J_{\lambda}$ of the ISRF at the desired locations.

In order to verify the validity of our method, we first considered a spherically symmetric test case that could be directly compared with a standalone {\sc Cloudy} simulation (a $3\times 10^4$\,K blackbody surrounded by a uniform ISM shell, simulating a B star in a H{\sc ii} region).
We find an excellent agreement between the ISRF predicted by both codes over the whole electromagnetic spectrum.

Then, we apply the model to our scientific problem, using {\sc Mixclask} to estimate the ISRF predicted by two different chemical evolution codes, {\sc Mulchem}~\citep{Molla+2023} and MMA~\citep{Millan-Irigoyen+20}, intended to represent a disk galaxy similar to the Milky Way.
Both CEMs provide surface densities of gas and stars across the Galaxy, but they differ in terms of their underlying physical assumptions as well as on their treatment of the ISM chemical composition (e.g. {\sc Mulchem} computes the evolution of many chemical species in the gas, whereas MMA tracks the formation and destruction of solid dust grains).
The ISRF returned by {\sc Mixclask} is in both cases broadly consistent with previous results reported in the literature about the spatial and spectral distribution of the radiation field in our Galaxy. This supports the computational validity of our algorithm in order to use with CEM and successfully predict the ISRF along with the main properties of the stars and interstellar medium.

However, the agreement between the different estimates in the literature, including our two models, is far from perfect, and the results presented in Section~\ref{sec_science} highlight several aspects that are worth a much deeper analysis.
On the one hand, our models seem to systematically underestimate the average intensity of the ISRF in the mid-IR regime, which may only be addressed by a thorough revision of the dust physics.
On the other hand, the radial dependence of the ISRF is far from being fully understood.
There are significant differences between the reported gradients, especially in the optical and ultraviolet range, where the number of models points is higher, and the spectrum of the ISRF close to the Galactic centre is still highly uncertain.
In particular, we would like to argue that the results presented here show the feasibility of the proposed method to study the significant systematic uncertainties that still remain in the estimation of the Galactic ISRF, especially far away from the Solar neighbourhood (both at the Galactic centre and the outer regions).

From a methodological point of view, the main advantage of the forward-modelling approach is that the observed discrepancies between different chemical evolution models may be traced to the physical processes and approximations that they consider, as well as their numerical implementation (e.g. spatial distribution of the infall rate, stellar IMF, nucleosynthetic yields, dust composition and size distribution, etc.), which are in turn subject to other independent observational constraints.
Conversely, enabling a CEM to predict the ISRF at different wavelengths helps to mitigate the degeneracy between the model's free parameters by broadening the range of empirical measurements available and allowing a more direct comparison between theory and data.
In reality, the observed emission as a function of wavelength is not independent from the star formation history, nor the chemical composition of the stars and the gas and dust in the different phases of the ISM.
A successful model must simultaneously fit all the observational data in order to provide a meaningful picture of the object under study.

Our two models of the Galactic ISRF may be easily incorporated into cosmic-ray propagation codes \citep[e.g.][]{DRAGON2}, the radiative heating and cooling prescription of \citet{Romero+21}, or any other practical application.
Once a model is fully calibrated, its parameters can be tuned to reproduce very different kinds of systems.
Although~\citet{PopescuTuffs13} presented a library of radiation fields of star-forming galaxies,
{\sc Mixclask} is well suited to predict the ISRF of \emph{any} galaxy that is well approximated by a certain chemical evolution model.
This includes a large variety of systems, from dwarf (irregular or spheroidal) to large (spiral or elliptical) galaxies, at any time along their evolution~\citep[see e.g.][]{Molla+2005, Gavilan+13}.
The ISRF features very different characteristics in these objects, and our code provides an efficient way to generate a self-consistent prediction that will be undoubtedly valuable in order to investigate any physical process that depends critically on the assumed ISRF, such as the evolution of supernova remnants or the propagation of cosmic rays.
We make our code freely available to the scientific community to tackle these problems and/or adapt it to model other classes of objects, such as individual stellar objects, stellar associations, galaxy clusters, etc.
As long as the user can provide the required information about the geometry of the system, the spectrum of the light sources, and the chemical composition of the surrounding medium, {\sc Mixclask} will be able to generate the corresponding ISRF.

\section*{Acknowledgments}

We would like to thank D. Gaggero, {as well as the anonymous referee}, for providing useful suggestions that have improved the quality of this article.
Financial support has been provided by grants PID2019-107408GB-C41 and PID2019-107408GB-C42 of the Spanish State Research Agency (AEI/10.13039/501100011033).
Calculations were performed with version 17.03 of {\sc Cloudy}, last described by~\citet{Cloudy17}; and version 9 of {\sc Skirt}, described in~\citet{Skirt9}.

\section*{Data availability}

Source code and input data regarding sections~\ref{sec_testing} and~\ref{sec_science} are found in \url{https://github.com/MarioRomeroC/Mixclask}.
Other data underlying this article will be shared upon reasonable request to the corresponding author.


\bibliographystyle{mnras}
\bibliography{references}

\appendix

\section{Skirt and Cloudy configuration}
\label{app_Technical}

In this appendix we provide technical details of how {\sc Skirt} and {\sc Cloudy} are configured when running {\sc Mixclask}.
It is assumed that the reader has some basic knowledge of both programs, and we refer to their respective documentation for further details.

\subsection{{\sc Skirt}}
\label{sapp_skirt}

We use the radiative transfer code {\sc Skirt} v9~\citep{Skirt9} to predict the mean intensity field in different positions of the spatial domain.
{\sc Mixclask} configures {\sc Skirt} input file in \skirtCommand{expert} mode.
All models shown in the present work discretise the computational domain into an axisymmetric spatial grid in the Local Universe (i.e. at redshift $z=0$).
A single {\sc Skirt} run is executed in each {\sc Mixclask} iteration, including all stellar and ISM regions defined by the user, where the trajectories of $N_{\gamma}$ photon packets (defined by the user) are used to estimate the ISRF based on the latest emissivities and opacities.

{\sc Skirt} makes a distinction between a \skirtCommand{source}, associated to the luminous component where the photons are generated, and a \skirtCommand{medium} that contains all the relevant information about the properties of the material through which the photons are travelling:

Under this scheme, all stellar \emph{and} ISM regions are considered \skirtCommand{sources} by {\sc Mixclask}, since they all emit radiation.
{\sc Skirt} needs the total luminosity $\lambda L_{\lambda}$ of each region (we restrict ourselves to isotropic and non-polarized emission) as a function of wavelength, which is given as input to {\sc Mixclask} for the stellar component, whereas it is iteratively computed by {\sc Cloudy} for the ISM based on the current estimate of the local ISRF.

On the other hand, a \skirtCommand{medium} in {\sc Skirt} is meant to handle the absorption and secondary emission by dust particles in the ISM.
However, we set the option \skirtCommand{dust extinction only} to disable the internal iterative scheme that {\sc Skirt} uses to determine the dust temperature and emissivity, because we call {\sc Cloudy} to carry out that task for all the ISM phases and components.
In order to input that information into {\sc Skirt}, each ISM region is considered both as a \skirtCommand{source} and a \skirtCommand{medium}, and we use the option of importing a dust mix by loading a text file that specifies the absorption and scattering due to the combined action of gas and dust.
This file includes, as a function of wavelength, the mass extinction coefficient $\kappa$, the albedo $\varpi$, and an scattering asymmetry parameter, which we set to $0$. 
The overall normalization of these terms is defined with the ISM mass $M$ of each region, which is one of the mandatory inputs in {\sc Mixclask}.

The output we request from {\sc Skirt} is provided by a \skirtCommand{probe} that stores the mean intensity of the radiation field at the centre of each ISM zone, required for the {\sc Cloudy} simulations, as well as any other set of locations specified by the user.

{\sc Mixclask} also takes two (optional) parameters from {\sc Skirt} to handle the numerical noise that is produced in each {\sc Skirt} simulation.
First one is to provide a file that stores a probability density distribution $p_{user}(\lambda)$, which is combined with a probability distribution representing the SED, $p_{SED}$, across the wavelength range specified by the user\footnote{See \url{https://skirt.ugent.be/skirt9/class_source.html} for more information}. 
The weight, $b$, of $p_{user}$ distribution is the second parameter, called \skirtCommand{wavelengthBias} in {\sc Mixclask} documentation.
This way, the final probability distribution in the range $(\lambda_0,\lambda_1)$ is:
\begin{equation}
    p(\lambda) = b p_{user}(\lambda) + (1-b)p_{SED}
\end{equation}
Probability distribution given by the user does not have to be normalized, as {\sc Skirt} does that internally.

\subsection{{\sc Cloudy}}
\label{ssec_cloudy}

We use the photoionisation code {\sc Cloudy} v17.03~\citep[last described in][]{Cloudy17} to characterize the emission and absorption of a region with gas and dust.
For each region, the radiation field given by {\sc Skirt} is set with the \cloudyCommand{table sed} command and its normalization with \cloudyCommand{nuf(nu) ... at ...}.
Values of both commands come from the results of the previous {\sc Skirt} run (see above).
We stop when the thickness of the cloud modelled in {\sc Cloudy} is the half of the width of each ISM region in {\sc Mixclask}, setting the command \cloudyCommand{Stop temperature off} to avoid another stop conditions.
No iterations are asked inside {\sc Cloudy} for the same reasons as {\sc Skirt}.

Chemical composition is treated with three levels of detail, although all of them requires to define $n_H$ with \cloudyCommand{hden}, and include \cloudyCommand{cosmic rays background} to include cosmic rays.
By default, if no additional input is given (e.g. metallicity, {\rm He} fraction, etc), {\sc Cloudy} uses \cloudyCommand{abundances gass} as chemical compositon~\citep{GASS}.
The user can give either {\sc He} fraction and metallicity or specific element fractions to change the default values to those desired.
One important detail about custom abundances in {\sc Cloudy} is that it asks for the number of particles relative to Hydrogen (i.e.:$N_i/N_{\rm H}$), while {\sc Mixclask} demands mass fractions ($X_i = M_i/M_{gas}$). The conversion between both is
\begin{equation}
    \frac{N_i}{N_{\rm H}} = \frac{X_i}{A_i X_H}
\end{equation}
where $A_i$ is the mass number of the element $i$.

Dust is included by default as grains and PAH separately, and can be tuned with $DTG$ and $q_{\rm PAH}$ in the input file.
Grains are added with \cloudyCommand{grains ism} scaled with $DTG \times (1-q_{\rm PAH})$, while PAH uses \cloudyCommand{grains pah}, scaled with $DTG \times q_{\rm PAH}$, paired with \cloudyCommand{set pah constant} to place them through the whole cloud.
For both dust types, we give the value relative to their default abundances, which are $6.6\times 10^{-3}$ and $2.6\times 10^{-5}$ for \texttt{grains ism} and \texttt{pah}, respectively.

We want as output ISM emission spectra and opacity.
For the emission, we ask {\sc Cloudy} to return the emitted continuum, given as $4\pi \lambda J_{\lambda}$, as a function of wavelength from the gas, which is one of the columns that the command \cloudyCommand{Save continuum} gives.
Opacities, on the other hand, are obtained from the optical depths, $\tau$, as a function of wavelength by using the \cloudyCommand{save optical depth} command.

\section{{\sc cloudy} input file for the HII region}
\label{app_cloudyControl}

Table~\ref{table_control} presents the input file needed to reproduce the {\sc Cloudy} control run of the H{\sc ii} region (section~\ref{sec_testing}):
{\sc Cloudy} considers this simulation as a \cloudyCommand{luminosity case}, and thus the output spectra are given as neutral luminosity, $\lambda L_{\lambda}$.
To convert it into a mean intensity, we follow section 2.6 of Hazy II (part of the {\sc Cloudy} documentation), and we divide $\lambda L_{\lambda}$ by $4\pi r^2$, where $r=40$\,pc corresponds to the outer radius of the spherical shell.

\begin{table}
    \begin{tabular}{l}
        Commands \\
        \hline
        \texttt{title cloudy control of section 3} \\
        \texttt{\#\# CHEMICAL COMPOSITION } \\
        \texttt{hden 0} \\
        \texttt{abundances gass} \\
        \texttt{grains ism} \\
        \texttt{grains pah} \\
        \texttt{set pah constant} \\
        \texttt{cosmic rays background} \\
        \texttt{\#\# RADIATION FIELD} \\
        \texttt{blackbody 3e4} \\
        \texttt{luminosity 38.28 range 0.1 to 1e9 nm} \\
        \texttt{\#\# OTHER OPTIONS} \\
        \texttt{stop temperature off} \\
        \texttt{iterate to convergence} \\
        \texttt{\#\# GEOMETRY} \\
        \texttt{radius 10.0 parsec linear} \\
        \texttt{stop thickness 30.0 parsec linear} \\
        \texttt{\#\# OUTPUTS} \\
        \texttt{save continuum 'spectra\_control.txt' last units nm } \\
        \hline
    \end{tabular}
    \caption{Input file to generate the {\sc Cloudy} solution of a H{\sc ii} region. $3e4$ and $38.28$ refers to the blackbody temperature (in K) and the decimal logarithm of luminosity (in erg/s), respectively.}
    \label{table_control}
\end{table}

\label{lastpage}
\end{document}